\documentclass[reqno,a4paper,12pt]{amsart}
\usepackage{amsmath,amssymb,amsthm,amscd,amsfonts,amsbsy,bm}
\usepackage[english]{babel}
\usepackage[mathscr]{eucal}

\font\sc=rsfs10 at 12pt

\numberwithin{equation}{section}

\renewcommand{\a}{\alpha}
\renewcommand{\b}{\beta}
\newcommand{\g}{\gamma}
\newcommand{\G}{\Gamma}
\renewcommand{\d}{\delta}
\newcommand{\D}{\Delta}
\newcommand{\e}{\epsilon}

\newcommand{\z}{\zeta}

\renewcommand{\k}{\kappa}

\renewcommand{\l}{\lambda}
\renewcommand{\L}{\Lambda}
\newcommand{\m}{\mu}
\newcommand{\n}{\nu}
\newcommand{\x}{\xi}
\newcommand{\X}{\Xi}
\renewcommand{\r}{\rho}
\newcommand{\s}{\sigma}

\renewcommand{\t}{\tau}

\newcommand{\p}{\psi}
\renewcommand{\P}{\Psi}

\newcommand{\C}{{\mathbb C}}
\newcommand{\R}{{\mathbb R}}

\newcommand{\ab}{{\mathbf a}}

\newcommand{\bb}{{\mathbf b}}

\newcommand{\Ab}{{\mathbf A}}

\newcommand{\Bb}{{\mathbf B}}

\newcommand{\Db}{{\mathbf D}}

\newcommand{\Mb}{{\mathbf M}}

\newcommand{\Rb}{{\mathbf R}}

\newcommand{\AF}{\mathfrak A}

\newcommand{\DF}{\mathfrak D}

\newcommand{\FF}{\mathfrak F}

\newcommand{\GF}{\mathfrak G}
\newcommand{\HF}{\mathfrak H}

\newcommand{\KF}{\mathfrak K}

\newcommand{\LF}{\mathfrak L}

\newcommand{\MF}{\mathfrak M}

\newcommand{\PF}{\mathfrak P}

\newcommand{\QF}{\mathfrak Q}

\newcommand{\RF}{\mathfrak R}

\newcommand{\SF}{\mathfrak S}
\newcommand{\tF}{\mathfrak t}
\newcommand{\TF}{\mathfrak T}
\newcommand{\wF}{\mathfrak w}

\newcommand{\VF}{\mathfrak V}

\newcommand{\Fc}{{\mathcal F}}
\newcommand{\Gc}{{\mathcal G}}
\newcommand{\Hc}{{\mathcal H}}

\newcommand{\Lc}{{\mathcal L}}

\newcommand{\Tc}{{\mathcal T}}
\newcommand{\Xc}{{\mathcal X}}
\newcommand{\Yc}{{\mathcal Y}}
\newcommand{\Zc}{{\mathcal Z}}

\newcommand{\Ls}{\sc\mbox{L}\hspace{1.0pt}}
\newcommand{\Ms}{\sc\mbox{M}\hspace{1.0pt}}

\newcommand{\diag}{{\rm diag}\,}
\newcommand{\curl}{{\rm curl}\,}

\newcommand{\sgn}{\hbox{{\rm sign}}\,}
\newcommand{\supp}{\hbox{{\rm supp}}\,}

\newcommand{\Bn}{\mathbf{B^\circ}}
\newcommand{\Q}{\QF}
\newcommand{\Qa}{\overline{\QF}{}}
\newcommand{\QFc}{\overline{\QF^\circ}{}}
\newcommand{\pd}{\partial} 
\newcommand{\pa}{\bar{\partial}}

\newcommand{\Rm}{R_-}
\newcommand{\Rp}{R_+}
\newcommand{\Cp}{{\rm Cap}\,}
\newcommand{\bl}{\bm{[}}
\newcommand{\br}{\bm{]}}
\newcommand{\bs}{\bb^{\sharp}}
\newcommand{\as}{\ab^{\sharp}}

\newcommand{\ps}{\psi^{\sharp}}

\newcommand{\codim}{\operatorname{codim\,}}

\newcommand{\Dom}{\operatorname{Dom\,}}

\newcommand{\dom}[1]{\DF({#1})}

\newtheorem{theorem}{Theorem}[section]
\newtheorem{proposition}[theorem]{Proposition}
\newtheorem{lemma}[theorem]{Lemma}
\newtheorem{corollary}[theorem]{Corollary}

\theoremstyle{definition}

\theoremstyle{remark}
\newtheorem{remark}[theorem]{Remark}

\begin{document}

\title[Perturbed Landau Hamiltonian]{On the Spectral
 Properties of the Perturbed Landau Hamiltonian}
 \dedicatory{Dedicated to our teacher, Professor Mikhail Solomyak\\ on the occasion of
his 75-th
 birthday}
\author[Rozenblum]{Grigori Rozenblum}
\address[G. Rozenblum]{Department of Mathematics \\
                        Chalmers University of Technology \\
                        and University of Gothenburg \\
                        Chalmers Tv\"argatan,3 , S-412 96
                         Gothenburg \\
                        Sweden}
\email{grigori@math.chalmers.se}
\author[Tashchiyan]{Grigory  Tashchiyan}
\address[G. Tashchiyan]{Department of Mathematics
\\                    St. Petersburg University
for Telecommunications\\
St.Petersburg, 198504, Russia}
\email{grigori.tashchiyan@gmail.com}

\begin{abstract}
The Landau Hamiltonian governing the behavior of a
quantum particle in dimension 2 in a constant magnetic
field is perturbed by a compactly supported magnetic
field and a similar electric field. We  describe how
the spectral subspaces change and how  the Landau
levels split under this perturbation.
\end{abstract}

\date{}

\maketitle

 \tableofcontents

\section{Introduction}
\label{intro}

The paper is devoted to the study of the spectrum of
the Schr\"odinger and Pauli operators in the plane,
with nonzero constant magnetic field perturbed by
smooth compactly supported magnetic and electric
fields.

 The Landau Hamiltonian describing the
motion of the quantum particle in two dimensions under
the influence  of the constant magnetic field is one
of the classical models in quantum physics. It was
introduced and studied in late 1920-s, see
\cite{LanLif}. It turned out that the system possesses
rather unusual spectral properties: the spectrum
consists of discrete eigenvalues lying at the points
of an arithmetic progression and all these eigenvalues
have infinite multiplicity. These eigenvalues are
traditionally called Landau levels (LL) and the
corresponding spectral subspaces are called Landau
subspaces.

A natural question arises, what happens with the
spectrum of the Landau Hamiltonian under
perturbations. If the perturbation is strong, say, it
is a Coulomb electric field, it turns out more
reasonable to consider the constant magnetic field as
a perturbation, and it leads to Zeeman effect. On the
other hand, one can consider weak perturbations of the
Landau Hamiltonian by a fast decaying or even
compactly supported electric or/and magnetic field. In
this case it is natural to expect that the Landau
levels will split, and the question consists in
describing this splitting as well as the behavior of
the spectral subspaces under the perturbation.

The case of the perturbation by an electric potential,
moderately, power-like decaying at infinity,  was
first studied by Raikov in \cite{Raikov1}. It was
shown that each LL splits into an infinite sequence of
eigenvalues, converging to the originating LL. The
rate of convergence of these eigenvalues is also
power-like and it is determined by the rate of decay
of the perturbation at infinity, the asymptotics being
described  by a semi-classical formula.  Advanced
methods of micro-local analysis were used in
\cite{Raikov1}.

The case of a fast decaying (or compactly supported)
electric potential was dealt with much later, in
\cite{RaiWar}, see also \cite{MelRoz}. It turned out
that the Landau levels still split, however in an
unusual, not a semi-classical way.  Most profoundly
this can be seen for a compactly supported
perturbation. The semi-classical considerations,
formally applied, would lead to a finite number of
eigenvalues splitting away from the LL. However, in
fact, not a finite but an infinite number of
eigenvalues split away, but they converge to the LL
extraordinarily fast, super-exponentially. Moreover,
the leading term in the asymptotics does not depend on
the perturbation at all, and the next term (as it was
found recently in \cite{FilPush}) is determined by the
geometry of the support of the perturbation. This
result can be also understood as justifying the first
order Raleigh-Schr\"odinger perturbative correction,
with properly chosen initial basis in the Landau
subspace. Methods of micro-local analysis do not work
here, and the results are obtained by a combinations
of methods of complex analysis and variational
considerations.

Considerably less is known about the perturbations of
the Landau Hamiltonian by a weak magnetic field. The
complication here has its roots in the fact that it is
not the magnetic field itself  but its potential that
enters in the quantum Hamiltonian. Unless the total
flux of the perturbing magnetic field vanishes, the
perturbation of the (Schr\"odinger or Pauli) operator
contains rather slowly decaying coefficients, so it is
 fairly strong  even for a compactly supported
perturbation of the field and it  even may be not
relatively compact if the perturbation goes to zero at
infinity  not sufficiently fast. Iwatsuka \cite{Iwats}
proved that invariance of the essential spectrum still
takes place, so Landau levels are the only possible
limit points of eigenvalues in the gaps between them.
After that, there were very few results in the
spectral analysis of this problem (apart from the
rather specific situation with the lowest LL). In the
paper \cite{Exner} the perturbation by a single
Aharonov-Bohm solenoid was considered. For the
standard self-adjoint extension, by means of
explicitly solving the eigenvalue problem, it was
established that the Landau levels are still intact,
the eigenspaces change in a controlled way, and a
finite number of eigenvalues split away  from each LL.
Hempel and  Levendorski in \cite{HemLev} and Besch in
\cite{Besch} investigated the behavior of the
eigenvalues in the gaps when a large parameter tending
to infinity stands in front of the perturbation. The
number of eigenvalues between the Landau levels grows
quasi-classically, as a power of the parameter, but
nothing was found out about the distribution of the
eigenvalues for a fixed value of the parameter.
However it was shown that one should not expect
monotonicity of the eigenvalues in this parameter,
even if the perturbing field has constant sign.

The aim of the present paper is to find out what
happens with the spectrum and spectral subspaces of
the Landau Hamiltonian  under smooth compactly
supported perturbations of the magnetic field,
 with additional perturbation by a compactly supported
  electric field (thus
both the case of the magnetic Schr\"odinger and of the
Pauli operators are taken care of). We show that the
Landau levels split superexponentially, and under
certain  positivity conditions find the asymptotics of
eigenvalues as they approach the LL. We find also a
rather exact approximation to the spectral subspaces
of the perturbed operator. It turns out that the
spectral subspaces are perturbed fairly strongly. This
effect is absent for the perturbation by the electric
field only: we show that in the latter case the
spectral subspaces change very weakly, in proper terms
(the behavior of spectral subspaces was not
investigated in \cite{RaiWar}, \cite{MelRoz}).

Our approach is based upon the study of approximate
creation and annihilation operators. The commutation
relations for these operators (the deformed Heisenberg
relations) enable us to construct approximate spectral
subspaces. It is known that for the unperturbed
operator the Landau subspaces are obtained by
repeatedly  applying the creation operator to the
explicitly described infinite-dimensional space of
zero modes of the Pauli operator. The Pauli operator
with the perturbed magnetic field  possesses an
infinite-dimensional space of zero modes as well,
moreover, this subspace  can also  be explicitly
described. It turns out that the application of the
same procedure for the perturbed operator, but using
the approximate creation operator instead, produces a
very good approximation for the spectral subspaces for
the perturbed operator.

In the papers \cite{RaiWar}, \cite{MelRoz},
\cite{FilPush}, the splitting of Landau levels under
the electric perturbation was associated with the
spectrum of certain Toeplitz-type operators acting in
the spectral subspaces. In our case, the study of the
spectrum of the perturbed operator is based upon  the
analysis of   Toeplitz-type operators in the space of
zero modes,  generalized in the sense that a
differential operator replaces the multiplication by a
function in the Toeplitz expression.

We prove the asymptotics of the eigenvalues split from
the Landau level  under a certain positivity condition
imposed on the \emph{effective} potential reflecting
the combined influence of the electric and magnetic
perturbations upon the spectrum. If this condition is
not fulfilled, we can only prove  the superexponential
estimate from above. The obstacle here is the lack of
knowledge of the spectrum of the Toeplitz-like
operators with indefinite weight in the space of zero
modes, which turns out to be a separate hard problem
in complex analysis.

In Section 2 we define our main operators and
establish a series of relations between them. Further,
in Section 3, we construct the approximate spectral subspaces.
 In Section 4 we prove that they form a very good approximation
  for  the actual spectral subspaces. As a
by-product, we show that the spectral subspaces
'almost' do not change if just an electric
perturbation is present. Section 5 is devoted to the
analysis of the Toeplitz-like operators. In Sect. 6 we
establish the estimates, and when possible, the
asymptotics in the Landau levels splitting. The last
section contains the proofs of several rather
technical formulas and estimates.

Different results involve different smoothness
conditions imposed on the perturbations. The upper
estimate for the eigenvalue splitting is proved for
$C^2$ compactly supported perturbations. On the other
hand, the lower estimates granting the eigenvalue
asymptotics, as well as the result on the approximate
spectral subspaces are proved for infinitely
differentiable perturbations. These latter smoothness
conditions can be replaced by some finite smoothness,
depending on the Landau level under consideration, but
we do not go into these details.

The conditions of the perturbations to have compact
support can be replaced by the requirements of
controlled decay at infinity. In that case the results
on the eigenvalue splitting and on the spectral
subspaces approximation  can also be obtained, with
the eigenvalue estimates and asymptotics as well as
the quality of spectral subspaces approximation
depending on the rate of decay. These resulst will be
published elsewhere.

The second author was partly supported by a grant from
the Royal Swedish Academy of Sciences, he also thanks
Chalmers University of Technology in Gothenburg for
hospitality.

\section{Magnetic Schr\"odinger  and Pauli operators}
\label{setting}

We will denote the points in the plane $\R^2$ by
$x=(x_1,x_2)$; it is convenient to identify $\R^2$
with $\C$ by setting $z=x_1+i x_2$. The Lebesgue
measure  will be  denoted by $dx$ and the derivatives
by $\pd_k=\pd_{x_k}$;  we set, as usual,
$\pa=(\pd_1+i\pd_2)/2,\; \pd=(\pd_1-i\pd_2)/2.$

The constant magnetic field will be denoted by $\Bn$;
we fix the sign, $\Bn>0$.  With this magnetic field we
associate the magnetic potential
$\Ab^{\!\circ}(x)=(A_1^\circ,A_2^\circ)=\frac{\Bn}{2}(-x_2,x_1)$.
Then the (unperturbed) magnetic Schr\"odinger operator
in $L_2(\R^2)$ is defined as
\begin{equation}\label{2:Schr.0}
    \HF^\circ=-(\nabla-i\Ab^{\!\circ})^2.
\end{equation}
The Pauli operator describing the motion of a
spin-${\frac12}$ particle acts in the space
$L_2(\R^2)^2$ of two-component vector functions, it
has the diagonal form,
$\PF^\circ=\diag(\PF^\circ_+,\PF^\circ_-)$, where
$\PF^\circ_\pm=\HF^\circ \pm \Bn$.

The spectrum of the Schr\"odinger operator was found
by Landau in 1928 (see \cite{LanLif}) by separation of
variables. However more conveniently  the spectral
structure of these operators is  described by means of
creation and annihilation operators introduced by Fock
\cite{Fock} and reflecting  the algebraical structure
of the problem (see, for example the detailed
exposition in~\cite{Folland} or \cite{BerShu}).
  Let us introduce the  complex magnetic potential
$A^\circ=A_1^\circ+iA_2^\circ$ and set
\begin{equation}\label{2:Creation.0}
    \Q^\circ=-2i\pa-{A^\circ},\; \QFc=-2i\pd-\overline{A^\circ}.
\end{equation}
These operators can be also expressed by means of the scalar
potential of the magnetic field, the function
\begin{equation}\label{2:Creation.00}
\Psi^\circ(z)=\frac{\Bn}4|z|^2\end{equation}
 solving the equation
$\D\Psi^\circ=\Bn$:
\begin{equation}\label{2:Creation.01}
\nonumber
 \QFc=-2i e^{\Psi^\circ}\pd
e^{-\Psi^\circ},\; \Q^\circ=-2i e^{-\Psi^\circ}\pa
e^{\Psi^\circ}.
\end{equation}

  The creation and annihilation
operators $\QFc ,\Q^\circ $ satisfy the following
basic relation
\begin{equation}\label{2:CommRel.0.1}
[\Q^\circ,\QFc]=2\Bn.
\end{equation}
This relation, meaning that the operators
$\Q^\circ,\QFc,2\Bn$ generate a  representation of the
Heisenberg algebra, together with the formulas
\begin{equation}\label{2:CommRel.0.2}
\PF^\circ_+=\Q^\circ\QFc, \PF^\circ_-=\QFc\Q^\circ,
\HF^\circ=\Q^\circ\QFc-\Bn=\QFc\Q^\circ+\Bn,
\end{equation}
enable one to describe the spectrum completely, and
the construction is presented in numerous physical and
mathematical sources. Since $\QFc={\Q^\circ}^*$, the
equation $\PF^\circ_-u=0$ is equivalent to
\begin{equation}
\nonumber
 {\Q^\circ}u=e^{-\Psi^\circ}\pa( e^{\Psi^\circ}u)=0.
\end{equation}
This means that  $f=e^{\Psi^\circ}u$ is an entire
analytical  function  in $\C$, such that after
multiplication by $e^{-\Psi^\circ}$ it belongs to
$L_2$. The space of such  functions $f$ is called Fock
or Segal-Bargmann space $\Fc$ (see \cite{Bargmann},
\cite{BerShu},
 \cite{Folland} for extensive discussion); when we need to specify the
 particular weight involved in the definition of $\Fc$
  we will use the notation $\Fc_\Bn$. So, the null
subspace of the operator $\PF^\circ_-$, i.e., its
spectral subspace corresponding to the eigenvalue
$\L_0=0$, is $\Lc_0=e^{-\Psi^\circ}\Fc$. After this,
the commutation relations \eqref{2:CommRel.0.1},
\eqref{2:CommRel.0.2} imply that $\L_1=2\Bn$ is the
lowest  eigenvalue of the operator
$\PF^\circ_+=\PF^\circ_-+2\Bn$, with the same spectral
subspace, and thus $\Lc_1=\QFc\Lc_0$ is the spectral
subspace of $\PF^\circ_-$ with eigenvalue $\L_1$ and
so on. So, the spectrum of $\PF^\circ_-$ consists of
eigenvalues, \emph{Landau levels} $\L_q=2q\Bn$,
$q=0,1,\dots$, and the spectra of $\HF^\circ$,
$\PF^\circ_+$ consist, respectively, of $\L_q+\Bn$ and
$\L_q+2\Bn$. The operators $\QFc, \Q^\circ$ act
between \emph{Landau subspaces} $\Lc_q=\QFc^q \Lc_0,
q=0,1,\dots$,
 \begin{equation}\label{2:CrAnn}\QFc:
\Lc_q\mapsto \Lc_{q+1},\; \Q^\circ :\Lc_q\mapsto
\Lc_{q-1},\; \Q^\circ:\Lc_0\mapsto 0,\end{equation}
 and are, up to
constant factors, isometries of Landau subspaces.

It is also convenient to introduce the spectral
projection $P_q^\circ$ of $\PF^\circ_-$ corresponding
to the eigenvalue $\L_q=2q\Bn$. The above reasoning
can be expressed as
\begin{equation}\label{2:Projections}
    P_q^\circ=C_q^{-1}\QFc^q P_0^\circ {\Q^\circ}^q,
    \; C_q=q!(2\Bn)^q.
\end{equation}

Now we introduce the perturbation $\bb\in
C_0^\infty(\R^2)$ of the magnetic field, and we set
$\Bb=\Bn+\bb$. Let $\psi$ be a scalar potential for
the field $\bb$, a solution of the equation
$\D\psi=\bb$. Of course, $\psi$ is defined up to a
harmonic summand, the choice of $\psi$ corresponds to
the choice of gauge. The magnetic potential
$\ab=(a_1,a_2)=(-\pd_2\psi,\pd_1\psi)$, $\curl
\ab=\bb$, is thus determined up to a gradient, and the
complete scalar and vector magnetic potentials are
\begin{equation}\label{2:potentials}
   \Psi=\Psi^{\circ}+\psi,\; \Ab=\Ab^{\circ}+\ab.
\end{equation}
The crucial observation here is that although $\bb$
has compact support, we should not expect the same, or
even a rapid decay, from the magnetic potential $\ab$,
unless the total flux of $\bb$, $\int \bb(x)dx$ is
zero. This can be seen in the easiest way by applying
Green's formula, saying that the circulation of the
vector field $\ab$ along the circle with radius $\Rb$
equals the flux of $\bb=\curl\ab$ through the disk
$|x|<\Rb$. Thus, if the total flux of $\bb$ is nonzero
and  $\Rb$ is taken large enough, so that the support
of $\bb$ lies in the disk, the circulation of $\ab$ is
bounded away from zero, and thus $\ab$ may not decay
faster than $C\Rb^{-1}$, at least in some directions.
This corresponds to  a logarithmical growth rate of
$\psi$.

We define the perturbed magnetic Schr\"odinger
operator as in \eqref{2:Schr.0}, with $\Ab^\circ$
replaced by $\Ab$:
\begin{equation}
\nonumber
    \HF=-(\nabla-i\Ab)^2,
\end{equation}
and the components of Pauli operator as
\begin{equation}
 \nonumber   \PF_\pm=\HF\pm\Bb.
\end{equation}

It is easy to observe that the difference between
$\HF$ and $\HF^{\circ}$ contains an operator of
multiplication by $\Ab^\circ\cdot\ab$, the latter
function does not decay at infinity (again unless the
total flux of $\bb$ vanishes), and thus looks like
being \emph{not} a relatively compact perturbation of
$\HF^{\circ}$. However, thanks to the special form of
this term, the perturbation is still relatively
compact if $\ab\to 0$ at infinity, as  was noticed by
Besch \cite{Besch}.

Now let us look  at the algebraic structure related
with the perturbed operators. The perturbed creation
and annihilation operators are defined similarly to
\eqref{2:Creation.0}:
\begin{equation}
\nonumber \Qa=-2i\pd-\overline{A},\; \Q=-2i\pa-A,
\end{equation}
where $A$ is the complex magnetic potential,
$A=A^\circ +(a_1+ia_2)$. The representation via the
scalar potential takes the form
\begin{equation}
\nonumber    \Qa=-2i e^{\Psi}\pd e^{-\Psi},\; \Q=-2i
e^{-\Psi}\pa e^{\Psi}.
\end{equation}

Most important for us are the commutation relations
for $\Qa,\Q$:
\begin{equation}\label{2:ComRel.1}
[\Q,\Qa]=2\Bb=2\Bn+2\bb,
\end{equation}
together with
\begin{gather}\label{2:ComRel.2}
\PF_+=\Q\Qa,\; \PF_-=\Qa\Q, \;\PF_+-\PF_-=2\Bb=2\Bn+2\bb,\\
\HF=\Q\Qa-\Bb=\Qa\Q+\Bb.\label{2:ComRel.3}
\end{gather}

The relations \eqref{2:ComRel.1}-\eqref{2:ComRel.3}
look similar to \eqref{2:CommRel.0.1},
\eqref{2:CommRel.0.2}, however the crucial difference
is that the new ones contain variable functions on the
right-hand side, do not represent the Heisenberg
algebra any more and thus cannot be used for the exact
 spectral analysis. The spectra of Schr\"odinger and
 Pauli operators
 do not  determine each other any more. The only
information that one can obtain immediately, is the
description of the lowest point of the spectrum of
$\PF_-$.  Since, again, $\Qa=\Q^*$, the equation
$\PF_-u=0$ is equivalent to $\Q u=0$, or
$\pa(\exp(\Psi) u)=0$. So the function $f=u\exp\Psi$
is an entire analytical  function such that
$u=\exp(-\Psi) f\in L_2$. The space of entire
functions with this property is, obviously,
infinite-dimensional, it contains at least all
polynomials of ${z}$, although it does not necessarily
coincide with Fock space, since $\exp(-\psi)$ may have
power growth (or decay). We denote this space by
$\Fc^\bb$.

The same reasoning, applied to $\PF_+$, shows that the
null subspace of $\Qa$, as well as the one  of $\PF_+$
is zero: it consists of anti-analytical functions
which belong to $L_2$ after the multiplication by the
growing exponent $\exp(\Psi)$, and there are no such
functions, except zero.

We denote the null-space of $\PF_-$, the space of
\emph{zero modes}, by $\Hc_0$. In \cite{RozShir}
rather general sufficient conditions were found for
infiniteness of dimension of $\Hc_0$; in particular
they even  do not require the magnetic field to be a
weak perturbation of the constant one.

As it follows from the  relative compactness of the
perturbation, by Weyl's theorem, the essential
spectrum of the perturbed operator $\PF_-$ consists of
the same Landau levels $\L_q$, and the eigenvalues in
the gaps and below $\L_0$ may only have $\L_q$ as
their limit points. This latter fact was established
much earlier by Iwatsuka \cite{Iwats}, and the
reasoning based upon the commutation relations remains
valid even in the case of a slowly decaying magnetic
perturbation, when the perturbation of the operator is
not relatively compact any more. Since our
considerations in the paper can be considered as an
extension of the ones of Iwatsuka, we reproduce the
reasoning from  \cite{Iwats}.

\begin{theorem}\label{2:TheoIwatsuka}(\cite{Iwats}) The
 essential spectrum
of the operator $\PF_-$ consists of the points
$\L_q,\; q=0,1,\dots$. The essential spectra of $\HF$
and $\PF_+$ consist, respectively, of the points
$\L_q+\Bn$ and $\L_q+2\Bn$.\end{theorem}

\begin{proof}
Note first that the operator of multiplication by
$\bb$ is relatively  compact with respect to $\HF$ and
$\PF_\pm$. Thus, by Weyl theorem, the operators
$\PF_+$ and $\PF_-+2\Bn=\PF_+-2\bb$ have the same
essential spectrum, i.e. the essential spectrum of
$\PF_+$ is just the one of $\PF_-$, shifted by $2\Bn$.
So, $\PF_+$ has no essential spectrum in $[0,2\Bn)$.
 On the other hand, the nonzero spectra of
$\PF_+$ and $\PF_-$ coincide, since these operators
are just  products of $\Qa$ and $\Q$ in the opposite
order. Therefore $\PF_-$ has no essential spectrum in
$(0,2\Bn)$, and $2\Bn$ belongs to the essential
spectrum. Again, this means that the interval
$(2\Bn,4\Bn)$ is free from the essential spectrum of
$\PF_+$ and $4\Bn$ belongs to it. By repeating this
reasoning, the complete description of the essential
spectrum is obtained.
\end{proof}

Now we add a perturbation by the electric potential.
Let $V(x)$ be a real valued function in
$C_0^\infty(\R^2)$. We introduce the operators
\begin{equation}\label{2:ElField}
    \HF(V)=\HF+V, \; \PF_\pm(V)=\PF_\pm+V.
\end{equation}
Since the operator of multiplication by $V$ is
relatively compact with respect to $\HF, \PF_\pm$, the
operators \eqref{2:ElField} have the same essential
spectra as the respective unperturbed ones. In this
paper we are going to study the behavior of the
eigenvalues of $\HF(V), \PF_\pm(V)$ near these Landau
levels.

\section{Approximate spectral subspaces}
\label{Subspaces3} In this section we construct the
approximate spectral subspaces of the operators
\eqref{2:ElField} under the condition $\bb\in
C_0^\infty$. To do this, we need some more properties
of the space $\Hc_0$ of zero modes of $\PF_-$. We must
keep in mind that the ambiguity in associating the
vector potential with the magnetic field leads to the
ambiguity in defining the magnetic operators
themselves. The gauge transformations of the form
$u\mapsto e^{i\Theta}u$ with  real functions
$\Theta(z)$ such that $e^{i\Theta}$ is smooth, realize
the unitary equivalence of magnetic operators
constructed by different choice of the potential.

As it was explained in Section 2, the space $\Hc_0$
consists of functions $u(z)$ having the form
$u(z)=e^{-\Psi(z)}f(z)$ where $f(z)$ is an analytical
entire function such that $e^{-\Psi}f\in L_2$, and
$\Psi(z)$ is the potential in \eqref{2:potentials}.
Note first that the choice of $\Psi$ does not affect
$\Hc_0$, up to a gauge transformation. In fact, let
$\Phi$ be a harmonic function in $\R^2$, so that
$\Psi'=\Psi+\Phi$ is also a potential for the magnetic
field $\Bb=\Bn+\bb$, $\D \Psi'=\Bb. $  Denote by
$\Theta$ the function harmonic conjugate to $\Phi$, so
that $F=\Phi+i\Theta$ is an entire analytical
function. This gives us the representation
$$u(z)=e^{-i\Theta(z)}e^{-\Psi'(z)}(e^{F(z)}f(z)),$$
so the multiplication by $e^{i\Theta(z)}$ is the gauge
transformation in question.

However the perturbation of the magnetic field leads
to the perturbation of the space of zero modes that
cannot be reduced to a gauge transformation. In fact,
if there were a real function $\Theta(z)$ such that
$e^{i\Theta}\Hc_0=\Lc_0$, this would mean that the
function $e^{-\psi+i\Theta}$ is analytical. However in
this case the function $-\psi$, as the logarithm of
the absolute value of an analytical function, must be
harmonic, which it is not.

Nevertheless there is a close relation between $\Hc_0$
and $\Lc_0$, not reduced to a gauge transform. We fix
the potential $\psi$, defining it as the logarithmic
potential of the field $b$,
\begin{equation}\label{3:Potential}
    \psi(x)=(2\pi)^{-1}\int \bb(y)\log|x-y| dy.
\end{equation}

 The behavior of the potential $\psi$ at infinity is
determined by  $\s$, the total flux of the perturbing field divided
by $2\pi$,
\begin{equation}
 \nonumber   \s=(2\pi)^{-1}\int \bb(x)dx.
\end{equation}

If $\s=0$, $\psi$ is a bounded function, as well as
$e^{\pm\psi}$. So, the multiplication by the bounded
function $m=e^{-\psi}$ maps $\Lc_0$ to $\Hc_0$. This
mapping is, obviously, invertible.

Now suppose that $\s>0$. Then the function $\psi(x)$
grows at infinity as $\s\log|x|$, $m(x)=e^{\psi(x)}$
grows as $|x|^\s$ and it is not bounded any more. The
function $e^{-\psi(x)}$ is bounded. Thus the
multiplication by $m^{-1}$ is a bounded operator from
$\Hc_0$ to $\Lc_0$. The inverse mapping is not
bounded, but we show that, nevertheless, it is well
defined on a certain subspace in $\Lc_0$.
 Denote by
$\n$ the smallest integer larger than $\s$. Fix $\n$
arbitrary different points $z_1,\dots,z_\n$ in $\C$.
Denote by $p(z)$ the monic polynomial with simple
zeros at the points $z_1,\dots,z_\n$ and by
$\tilde\Lc_0$ the space of functions in $\Lc_0$ having
zeros at these points. Since the evaluation at a point
is a bounded functional in the Fock space,
$\tilde\Lc_0$ is a closed subspace in $\Lc_0$, with
codimension $\n$. Now consider the space
$\Lc^\n_0={p(z)}^{-1}\tilde\Lc_0$. By our
construction, this space is still a part of $\Lc_0$
(however is not a closed subspace any more). Since at
infinity we have $e^{\psi(z)}=O(p(z)^{-1})$, the
operator of multiplication by $e^{\psi(z)}$ maps
$\Lc^\n_0$ to $\Hc_0$. In the case of a negative $\s$,
the spaces $\Hc_0$ and $\Lc_0$ change places in the
above reasoning.

The space $\Hc_0$ is closely related to the spaces
$\FF^\r$ introduced in \cite{Bargmann}, Section 3,
consisting of entire analytical functions $f(z)$ that
belong to $L_2$ after being multiplied by
$\theta_\r(z)=(1+|z|^2)^{-\r/2}\exp(-(\Bn/4)|z|^2)$.
Due to \eqref{3:Potential}, the function
$\x(z)=\psi(z)-\frac{\s}2\log(1+|z|^2)$ is bounded, so
we can write
$$e^{-\Psi(z)}=\theta_{-\s}(z)e^{-\x(z)}.$$
Therefore our space $\Hc_0$ can be obtained from
Bargmann's space $\theta_{-\s}(z)\FF^{\s}$ by
multiplication by the bounded and boundedly invertible
function $e^{-\x(z)}$. This means that many properties
of this Bargmann space are inherited by our space
$\Hc_0$. What is important for us at the moment is the
fact established in \cite{Bargmann}, Section~3.6, that
the functions of the form $\theta_\s(z){z}^n$ form a
basis in $\theta_\s(z)\FF^{\s}$. Therefore this space
possesses a dense set consisting of rapidly decaying
functions (having the form $\theta_{-\s}(z) p({z})$,
with  polynomial $p$), and thus our null subspace
$\Hc_0$ possesses a dense subspace of rapidly decaying
functions as well.

Let $\d_q=(\L_q-\g,\L_q+\g), q=0,1,2,\dots$, be
intervals of the same size on the real axis, centered
at the Landau levels $\L_q=2q\Bn$, and not containing
the neighboring Landau levels $\L_{q\pm1}$.  We choose
the size of $\d_q$ in such way that  neither of these
intervals has the eigenvalues of $\PF_-$ at  its
endpoints. Moreover, since the lowest LL $\L_0=0$ is
an isolated point of the spectrum of $\PF_-$, we can
choose the size of the intervals in such way that
$\d_0$ contains only this point of spectrum.  We
denote by $\Hc_q$ the spectral subspace of $\PF_-$
corresponding to the interval $\d_q$ and by $P_q$ the
corresponding spectral projection. Since, by
Theorem~2.1, the spectrum of $\PF_-$ is discrete
between Landau levels, the change of $\d_q$ leads only
to a finite-rank perturbation of $P_q$. As usual, the
spectral projection $P_q$ can be be expressed by means
of the integration of the resolvent of $\PF_-$ along a
closed  contour $\G_q$ in the complex plane, not
passing through the eigenvalues of $\PF_-$ and
containing inside only those eigenvalues that lie in
$\d_q$. Again, using the discreteness of the spectrum
of $\PF_-$ between the Landau levels, we can choose
these contours in such way that all they are obtained
from $\G_0$  by the shift along the real axis in the
complex spectral plane, $\G_q=\G_0+2q\Bn$.

Now we are going to establish several properties of
the subspaces $\Lc_q$, projections $P_q$ and some
related operators.

First, note the simple fact following directly from
the spectral theorem.

\begin{proposition}\label{2:PropUnpertSubspace}For any
$q=0,1,\dots,$ and any polynomial $p(\l)$ the operator
$p(\PF_-) P_q$ is bounded, moreover the operator
$(p(\PF_-)-p(\L_q))P_q$ is compact.
\end{proposition}

In fact, by the spectral theorem the nonzero spectrum
of the operator $p(\PF_-)P_q$ consists of the points
$p(\l_j)$ where $\l_j$ are all points of spectrum of
$\PF_-$ in $\d_q$ and thus  all $p(\l_j)$  live in a
bounded interval. Moreover, $p(\l_j)$ may only have
$p(\L_q)$ as their limit point, by
Theorem~\ref{2:TheoIwatsuka}.

The following lemma  will
enable us later to establish a much stronger
compactness property.
\begin{lemma}\label{2:LemmaBound} Let each of
 $T_j,\; j=1,\dots,N$ be one of operators $\Q$ or
$\Qa$. Then for some constants $C, C'$, for any $u$ in
the domain of the operator $\PF_-^N$,
\begin{equation}\label{2:LemmaBound.1}
   \|T_1T_2\dots T_N u\|^2\le C(\PF_-^Nu,u)+C'\|u\|^2.
\end{equation}
\end{lemma}
The proof of Lemma~\ref{2:LemmaBound} can be found in
Section~\ref{technical}.

Now we can establish the compactness property.
\begin{proposition}\label{2:PropComp} Let
$T_1,\dots,T_N$ be a collection of operators, each
being $\Qa$ or $\Q$, and let $h_j$, $j=0,\dots N$ be
functions with all derivatives bounded, $\Tc=h_0T_1h_1
\dots T_N h_N$.
 Then for any
$q$ and for any polynomial $p(\l)$ the operators
$\Tc(p(\PF_-)-p(\L_q))P_q \Tc$
 are compact.\end{proposition}
\begin{proof}
By commuting functions $h_j$ and operators
$T_j$ (moving all functions to the left), we can transform
 the left operator $\Tc$ to the sum of
terms having the form $\tilde{h}_\k \Tc_\k$ where
$\tilde{h}_\k$ are bounded functions and $\Tc_\k$ is a
product of no more than $N$ operators $\Qa, \Q$.
Similarly, in the copy of $\Tc$ that stands to the
right from $P_q$, we move all functions to the utmost
right positions, to get the representation of $\Tc$ as
a sum of terms $\Tc_\varkappa\tilde{h}_\varkappa$,
being the product of a bounded smooth function
$\tilde{h}_\varkappa$ and no more than $N$ operators
$\Qa, \Q$.

For each of the terms arising in this way in the
decomposition of $\Tc(p(\PF_-)-p(\L_q))P_q \Tc$, we
can write
\begin{gather}\label{2:ProComp.1}
\tilde{h}_\k\Tc_\k (p(\PF_-)-p(\L_q))P_q
\Tc_\varkappa\tilde{h}_\varkappa = \bl\tilde{h}_\k
\Tc_\k(\PF_-^N+1)^{-1}\br\times \\\notag\bl(\PF_-^N+1)
(p(\PF_-)-p(\L_q))P_q(\PF_-^N+1)\br
\bl(\PF_-^N+1)^{-1}\Tc_\varkappa\tilde{h}_\varkappa\br.
\end{gather}

In \eqref{2:ProComp.1}, the first factor in brackets
is bounded by Lemma~\ref{2:LemmaBound}, and the middle
factor is compact by
Proposition~\ref{2:PropUnpertSubspace}. The last
factor in brackets is also bounded,
 by Lemma~\ref{2:LemmaBound} applied to the adjoint operator.
  \end{proof}

The result we have just proved indicates that,
probably, the eigenvalues of the operator $\PF_-$,
lying near $\L_q$, must converge to $\L_q$ rather
rapidly. This will be, in fact, established later on.

Now we are going to describe the main construction of
the paper, the approximate spectral subspaces of the
perturbed operator. Note, first of all, that it is
sufficient to consider the operator $\PF_-(V)$. In
fact, due to the commutation relations, we have
$$\HF(V)=\PF_-(V+\bb)+\Bn,
\PF_+(V)=\PF_-(V+2\bb)+2\Bn,$$ and thus the operators
differ from $\PF_-$ by a shift and  by the electric
type perturbations $V+\bb$, $V+2\bb$. We start by
finding approximate spectral subspaces of $\PF_-$.
Adding $V$ will be an easier task after this.

The subspaces approximating $\Hc_q$ will be defined in
the following way:
\begin{equation}\label{3:Subspaces}
    \Gc_0 =\Hc_0,\; \Gc_q=\Qa^q\Gc_0,\; q=1,2,\dots.
\end{equation}
So we mimic  the construction of the eigenspaces of
the unperturbed Landau Hamiltonian, see
\eqref{2:CrAnn}. Our approximate spectral subspaces
are obtained by applying the creation operators to the
space of zero modes.

Of course, since we apply the unbounded  operator
$\Qa$, we must show that all the time we do not leave
the space $L_2$,
 and moreover, we have to prove that the subspaces
$\Gc_q$ are closed. Both these properties, as well as
some other results of the paper will be based upon the
 important Proposition~\ref{3:basicPropositionBP} to follow.
 But first we need a definition.

Many times in the paper we will encounter  operators
that are products of several copies of $\Q,\Qa$,
functions $\Bb, \bb, U$ (for some smooth function $U$)
and their derivatives. Let us assign the weight $2$ to
$\Bn,\Bb,\bb,U$, the weight $l+2$ to the order $l$
derivatives of $\bb,U$ and the weight $1$ to $\Q$ and
to $\Qa$. For a product of operators and functions the
weights of factors are added.

\begin{proposition}\label{3:basicPropositionBP}
Let  $q>0$.\begin{enumerate}
 \item There exists a
function $\Zc_q[\bb]\in C_0^\infty$ depending only on
$q$, $\Bn$, and $\bb$ such that for any $u\in\Hc_0$,
\begin{equation}\label{3:BP:equation}
    \|\Qa^q u\|^2=C_q \|u\|^2+(\Zc_q[\bb] u,u),\; C_q=q!(2\Bn)^q.
\end{equation}
The function $\Zc_q[\bb]$ is  a polynomial of $\bb$
and its derivatives up to the order $2q-2$ with
coefficients depending on $\Bn$. All summands in
$\Zc_q[\bb]$ have weight $2q$. The term with the
highest derivative equals $2\D^q\bb$; the term linear
in $\bb$ and not containing derivatives equals
$C'_q\Bn^{q-1}\bb$, $C'_q=2^q q! q$.
 \item There exists a function $\Xc_q[\bb,U]\in C_0^\infty$ depending
 only on $q$, $\Bn$,  $\bb$, and $U$ such that for any
$u\in\Hc_0$.
\begin{equation}\label{3:BP:equation.V}
  (U\Qa^q u,\Qa^q u)= (\Xc_q[\bb,U] u,u).
\end{equation}
The function $\Xc_q[\bb,U]$ is expressible as an order
$2q$ linear differential operator acting on $U$, with
coefficients depending polynomially on $\bb,$ its
derivatives, and $\Bn$. All the terms in
$\Xc_q[\bb,U]$ have weight $2q+2$. The  term with the
highest derivative of $U$ is $\D^q U$. The term with
zero order derivative of $U$ is $C_q U+ \Zc_q[\bb] U$.
 \item There exists a function $\Yc_q[\bb,U]\in
C_0^\infty$ depending
 only on $q$, $\Bn$,  $\bb$, and $U$ such that for any
$u\in\Hc_0$.
\begin{equation}\label{3:BP:equation.VQ}
 (U\Qa^q u,\Qa^{q+1} u)= (\Yc_q[\bb,U] u,u).
\end{equation}
The function $\Yc_q[\bb,U]$ is expressible as an order
$2q+1$ linear  differential operator acting on $U$,
with coefficients depending polynomially on $\bb,$ its
derivatives, and $\Bn$. All the terms in
$\Yc_q[\bb,U]$ have weight $2q+3$. The  term with the
highest derivative of $U$ is $(-2i)\D^q \pa U$. The
term with zero order derivative of $U$ is  $C_q U+
\Zc_q[\bb] U$.
\end{enumerate}
\end{proposition}

\begin{proof} First, in order to make the main mechanism
 of the proof more clear,  we derive the relation
\eqref{3:BP:equation} for $q=1$ and $q=2$. It is
sufficient to prove \eqref{3:BP:equation} for
$u=\exp(-\Psi){p(z)}$ where $p(z)$ is a polynomial:
such functions are dense in $\Hc_0$, and then we can
pass to general $u$ by the closure. In the process of
our, rather lengthy, calculations we will repeatedly
use the commutation relations \eqref{2:ComRel.1} as
well as
\begin{equation}\label{3:BP:0}
    [\Q, h] =-2i\pa h, [\Qa,h]=-2i\pd h,\mathrm{ for \ } h\in
    C^\infty.
\end{equation}
    So, let us consider the
expression $\|\Qa u\|^2$, $u\in \Hc_0$. We have
\begin{gather}\label{3:BP:1}
\|\Qa u\|^2=(\Qa u, \Qa u)=(\Q\Qa
u,u)\\ \notag=(2\Bn+2\bb+\Qa\Q u,u)=2\Bn\|u\|^2+(2\bb u,u)
\end{gather}
(on the last step we used that $\Q u=0$). Thus we have
$\Zc_1[\bb]=2\bb$.

The case $q=2$ requires some more calculations. We
start with
\begin{equation}
\nonumber\|\Qa^2 u\|^2=(\Q^2\Qa^2u,u).
\end{equation}
Then we are going to commute $\Q$
 and $\Qa$ so that only $\Q$, and not $\Qa$ would act on
 $u$, so we will move all copies of $\Qa$ to the left and
 all copies of $\Q$ to the right:
 \begin{gather}\label{3:BP:3}
(\Q^2\Qa^2u,u)=(\Q(2\Bn+2\bb+\Qa \Q )\Qa
u,u)=2\Bn\|\Qa u\|^2 \\ \notag +2(\Q\bb\Qa
u,u)+(\Q\Qa\Q\Qa u,u).
\end{gather}
The first term in \eqref{3:BP:3} has already the
 form we aim for, by \eqref{3:BP:1}. In the second term we
commute $\Q$ and $\bb$ using \eqref{3:BP:0}:
\begin{equation}\label{3:BP:4}
2(\Q\bb\Qa u,u)=2(\bb\Q\Qa u,u)+2(-2i(\pa \bb)\Qa
u,u).
\end{equation}
In the first term in \eqref{3:BP:4} we commute $\Q$
and $\Qa$, and in the second one commute $-2i(\pa
\bb)$ and $ \Qa$:
\begin{gather}
\nonumber
2(\Q\bb\Qa u,u)= 4(\bb(\Bn+\bb) u,u)+\\
\notag 2(\bb\Qa\Q u,u)
      + 2(\Qa(-2i\pa\bb) u,u) +(8(\pd\pa \bb) u,u).
\end{gather}
Since $\Q u=0$, we arrive at
\begin{equation}
\nonumber 2(\Q\bb\Qa u,u)=4(\bb(\Bn+\bb) u,u)+2((\D
\bb) u,u).
\end{equation}
In the last term in \eqref{3:BP:3} we commute the
first pair  $\Q$ and $\Qa$ and then the second one:
\begin{gather}\label{3:BP:7}
(\Q\Qa\Q\Qa u,u)=(\Qa\Q^2\Qa u,u)+2\Bn(\Q\Qa
u,u)+2(\bb \Q\Qa u,u)\\\notag =2\Bn(\Qa u,\Qa
u)+4(\bb(\Bn+\bb)u,u)+2(\bb\Qa\Q u,u).
\end{gather}
On the right-hand side of \eqref{3:BP:7}, the first
term is again calculated by \eqref{3:BP:1}, the second
term has the required form, and the last one vanishes.
Now we collect all the transformed terms in
\eqref{3:BP:3} and obtain
\begin{gather}
\nonumber \|\Qa^2u\|^2=2\Bn(2\Bn\|u\|^2+(2\bb
u,u))+4(\bb(\Bn+\bb) u,u)+\\\notag 2(\D \bb
u,u)+2\Bn(2\Bn\|u\|^2+(2\bb u,u))+4(\bb(\Bn+\bb)u,u)\\
\notag=2(2\Bn)^2\|u\|^2 +(\Zc_2[\bb]u,u).
\end{gather}
with \begin{equation}
\nonumber    \Zc_2[\bb]=8\bb^2+12\Bn\bb+2\D\bb.
\end{equation}

Now we can formally describe how the procedure goes in
the general case. In the expression $\Q^q\Qa^q$ we
start commuting $\Q$ and $\Qa$, as well as these
operators with the functions arising by the previous
commutings. All the time we keep moving $\Q$ to the
right and $\Qa$ to the left. As soon as $\Q$ reaches
the utmost right position or $\Qa$ reaches the utmost
left position, the corresponding term vanishes. We do
this until no operators $\Q,\Qa$ are left, and only a
function $ C_q +\Zc_q[\bb]$ remains. To find $C_q$ and
describe the structure of $\Zc_q[\bb]$, we need only
 to do some bookkeeping
of the terms arising when the operators and/or
functions are commuted.   The formulas
\eqref{2:ComRel.1}, \eqref{3:BP:0} show that the
weight is preserved under commutation. So the constant
term $C_q$ has to be proportional to $\Bn^q$, and it
can arise only when the operators $\Q,\Qa$ are
commuted, not the operators with functions. The number
of such commutings needed to re-arrange all terms
equals $q!$.  The term linear in $\bb$ and not
containing derivatives results from the same procedure
and has the form $C'_q\Bn^{q-1} \bb$, $C'_q=2^nq!q$.
The function $\Zc_q[\bb]$ arising as a result of
moving all $\Q$-s to the right and all $\Qa$-s to the
left will be a polynomial of $\bb$ and its derivatives
with weight $2q$. The highest derivative of $\bb$
present in $\Zc_q[\bb]$ will thus be of order  $2q-2$,
and this term will be linear in $\bb$. Such term
arises when $2\Bb$ appears after the first commutation
of $\Qa,\Q$ in $\Q^q\Qa^q$,
\begin{equation}
\nonumber
\Q^q\Qa^q=\Q^{q-1}\Qa\Q\Qa^q+\Q^{q-1}(2\Bn+2\bb)\Q^{q-1},
\end{equation}
and then $q-1$ copies of $\Q$ pass $2\bb$ on their way
to the right and $q-1$ copies of $\Qa$ pass it on
their way to the left, each such pair producing the
Laplacian acting on $\bb$.

The above reasoning takes care of
\eqref{3:BP:equation}. Parts 2) and 3) of the
proposition are proved in the same way, with a similar
bookkeeping in order to determine the structure of the
resulting functions. We just demonstrate how Part 2)
works for $q=1$:\begin{gather}
\nonumber    (\Q U\Qa u,u)= (\Q\Qa U
u,u)+(\Q[U,\Qa]u,u)=  (\Qa\Q U
    u,u)(=0)+\\ \notag(2\Bb Uu,u)+([U,\Qa]\Q
    u,u)(=0)+([\Q,[U,\Qa]u,u)=(\Xc_1[\bb,U]u,u);
\end{gather}
$\Xc_1[\bb,U]=2\Bb U+\D U$.
\end{proof}

When applying  Proposition~\ref{3:basicPropositionBP}
and similar results, we need  a certain compactness
property. Such facts were used persistently in
\cite{RaiWar}, \cite{MelRoz}, \cite{FilPush}, but for
the case of a constant magnetic field only.
\begin{lemma}\label{3:LemComp}
Let $W(x)$ be a bounded function with compact support.
Then for any differential operator $\LF$ with bounded
coefficients the quadratic form
 $$\wF[u]=\int W(x)| \LF u|^2dx$$
 is compact in the space $\Hc_0$.\end{lemma}
 \begin{remark}\label{3:remark1}
 In section 5 we will find much more detailed
 information about such forms. However the proof we give here is
 also useful. We return to this approach later.\end{remark}
 \begin{proof} Let $u=e^{-\Psi}f$ be a function in
 $\Hc_0$, so $f(z)$ is an analytical function.

 Take some $\Rb$ such that the support of the function
 $W$ lies inside the circle $C_\Rb$ with radius $\Rb$
 centered in the origin. For each
 $r\in(\Rb,2\Rb)$ we write the Cauchy  representation
 for $f$:
 \begin{equation}\label{3:IntegralRepr1}
    f(z)=(2\pi i)^{-1}
    \int_{C_r}f(\z){(z-\z)}^{-1}
    d{\z}.
\end{equation}
Then for some fixed function $\x(r)\in
C_0^\infty(\Rb,2\Rb)$, $\int \x(r)dr =1$, we multiply
\eqref{3:IntegralRepr1} by $\x(r)$ and integrate in
$r$ from $\Rb$ to $2\Rb$. This gives the integral
representation of $f(x),\; |x|<\Rb$, in the form
$f(x)=$ $\int_{\Rb<|y|<2\Rb} K(x,y)f(y) dy$, with
smooth bounded kernel $K(x,y)$. After applying $\LF$
in $x$ variable, we obtain a similar integral
representation for $\LF u=\LF(e^{-\Psi}f)$:
\begin{equation}
\nonumber
    \LF u(x)=\int_{\Rb<|y|<2\Rb}
    e^{-\Psi(x)}K^{\LF}(x,y)e^{\Psi(y)}u(y)dy=(\KF^{\LF}u)(x).
\end{equation}
The integral operator $|W|^{1/2}\KF^{\LF}$ has a
bounded kernel with compact support and therefore
 is compact in
$L_2$, and thus the quadratic form $\wF[u]$ can be
written as $\wF[u]=(\sgn W |W|^{1/2}\KF^{\LF}u,$ $
|W|^{1/2}\KF^\LF u)$ and therefore it is
compact.\end{proof}

Now we are able to justify our construction of the
spaces $\Gc_q$.

\begin{proposition}\label{3:SubspacesGood} The sets
$\Gc_q$ defined in \eqref{3:Subspaces} are closed
subspaces in $L_2$.\end{proposition}
\begin{proof} The fact that $\Gc_q\subset L_2$ follows directly
from Proposition~\ref{3:basicPropositionBP}. Next, the
relation \eqref{3:BP:equation} can be written as
\begin{equation}\label{3:Subspaces:1}
    (P_0 \QF^q \Qa^q u,u)=C_q(u,u)+(P_0\Zc_q[\bb]
    u,u);\; u=\P_0 u\in\Hc_0
\end{equation}
(recall that $P_0$ is the orthogonal projection onto
$\Hc_0$, $P_0 u=u$). In the second term in
\eqref{3:Subspaces:1}, by Lemma~\ref{3:LemComp}, the
operator $P_0\Zc_q[\bb]$ is compact in $\Hc_0$, and
therefore we can understand \eqref{3:Subspaces:1} as
showing that the operator $C_q^{-1} P_0 \Q^q $ is a
left parametrix for $\Qa^q:\Hc_0\to L_2$. This implies
that the range of $\Qa^q$ is closed.\end{proof}

The null space of $\Qa$ and therefore of $\Qa^q$ is
zero. Consider the operator $\Qa^q$ as acting from
$\Gc_0=\Hc_0$ to $\Gc_q$. This is a bounded invertible
operator, therefore the inverse, that we denote by
$\Qa^{-q}$, is a bounded operator from $\Gc_q$ to
$\Gc_0$.  This inverse operator is a compact
perturbation of $\Q^q$.

We will use Proposition~\ref{3:basicPropositionBP} and
Lemma~\ref{3:LemComp} many times throughout the paper.
An immediate useful corollary of the latter lemma is the
following.
\begin{corollary}\label{3:Corollary} Under the
conditions of Proposition~\ref{3:basicPropositionBP},
for any $\e>0$ there exists a subspace $\Ms\subset
\Hc_0$ such that
 \begin{equation}
 \nonumber
\|\Qa^q v\|^2\ge (1-\e)C_q\|v\|^2, \; v\in\Ms,
\end{equation}
and the codimension of $\Ms$ in $\Hc_0$ is finite.
\end{corollary}
\section{Approximate spectral projections}\label{Projections4}

The aim of this section is to prove that the subspaces
$\Gc_q$ are very good approximations to the actual
spectral subspaces $\Hc_q$ of the operator $\PF_-$, as
well as to the spectral subspaces of $\PF_-(V)$.
Closeness of subspaces will be measured by closeness
of orthogonal projections onto them. Recall that the
projection onto $\Hc_q$ is denoted by $P_q$. Let $Q_q$
be the orthogonal projection onto $\Gc_q$.
\begin{theorem}\label{4:ThmCloseness}The projections
$P_q$ and $Q_q$ are \emph{close}: for any $N$, and any
collection of the operators $T_j,\; j=1,\dots,N$, each
 of $T_j$ being $\Q$ or $\Qa$,
the operator $\Tc(P_q-Q_q)\Tc$, is compact,
$\Tc=T_1T_2\dots T_N$.\end{theorem}

In justifying the theorem, we need two technical
lemmas which will be proved in
Section~\ref{technical}.

Both lemmas concern the operator properties of products
 of many copies of the
resolvents of $\PF_+$ and $\PF_-$, the creation and
 annihilation operators $\Qa$ and $\Q$,   functions
$h_j$ with all derivatives bounded, and, possibly, the
 spectral projection $P_q$ onto $\Hc_q$.
In such product, we assign order $1$ to $\Qa$ and
$\Q$, order $-2$ to the resolvent, order $0$ to
functions and projections. The order of the product is
defined as the sum of orders of factors.

\begin{lemma}\label{4:LemmaComp1}Let the operator $\AF$ be
 the product
  of
creation, annihilation operators, resolvents, and
functions $h_j$, have negative order, and let at least
one of the functions $h_j$ have compact support. Then
$\AF$ is compact.\end{lemma}
\begin{lemma}\label{4:LemmaComp2} Let the operator $\AF$ be
the product of creation, annihilation operators,
resolvents, functions and a spectral projection $P_q$.
Then $\AF$ is bounded. If, additionally,
  at least one of the functions has compact
support then $\AF$ is compact.\end{lemma}

The proof of the theorem  goes in the following way.
We  construct an intermediate operator $\SF_q$ with
range in $\Gc_q$ and prove that $\SF_q$ is close both
to $P_q$ and $Q_q$.

For $q>0$ we define the operator $\SF_q$ by the
relation
\begin{equation}\label{4:ApproxProj}
    \SF_q=C_q^{-1}\Qa^q P_0 \Q^q,\; C_q=q!(2\Bn)^q.
\end{equation}
So, our expression for the approximate spectral
projection is just a natural modification of the exact
formula \eqref{2:Projections} for the unperturbed
operator. Equivalently, the operator $\SF_q$ can be
described by the formula
\begin{equation}
\nonumber
 \SF_q=C_q^{-1} \GF \GF^*,\qquad \GF=\Qa^qP_0;
\end{equation}
in
Proposition~\ref{3:basicPropositionBP} this operator is shown to be bounded.

The proof of Theorem~\ref{4:ThmCloseness} will consist
of two parts, showing that $\SF_q$ is close to $P_q$
and showing that it is close to $Q_q.$

\begin{proposition}\label{4:Prop:SclosetoP} The
operator $\SF_q$ is close to the projection $P_q$.
\end{proposition}
\begin{proof} We give the detailed proof for $q=1$ first, and
 and
 then explain the natural changes one should make for
 larger $q$.

 We denote by $\Rm(\z)$ the
resolvent of the operator $\PF_-$; we also need the
resolvent $\Rp(\z)$ of the operator $\PF_+$. The
projection $P_0$ can be expressed via Riesz integral
\begin{equation}
\nonumber P_0=(2\pi i)^{-1}\int_{\G_0}\Rm(\z)d\z,
\end{equation}
where $\G_0$ is the closed curve, encircling the zero
point $\L_0$,  defined, together with curves $\G_q$,
in Sect.3. We are going to transform the expression
for the resolvent $\Rm(\z)$ using the commutation
relations \eqref{2:ComRel.1}, \eqref{2:ComRel.2},
\eqref{2:ComRel.3}, and \eqref{3:BP:0}. After this,
the crucial observation is that the integral of
$R_\pm^k,\, k\ge2$ along a closed contour vanishes.
This enables us to dispose of terms that are not
compact before integration.

 We start by
writing
\begin{equation}
\nonumber \Rm(\z)=(\PF_--\z)^{-1}=(\PF_+ -2\Bn -2
\bb-\z)^{-1}.
\end{equation}
Next we use the resolvent equation to get
\begin{gather}\label{4:TransRes2}
\Rm(\z)=(\PF_+ -2\Bn -\z)^{-1}-\\ \notag (\PF_--\z)^{-1}(2\bb)
(\PF_+ -2\Bn -\z)^{-1}=\Rp(2\Bn+\z)-H(\z)
\end{gather}
We multiply  \eqref{4:TransRes2} by
 $\Qa$ from the left and by $\Q$ from the right, as \eqref{4:ApproxProj}
  requires. For the first term  we use that
\begin{gather}\label{4:TransRes2.1}\Qa \Rp(2\Bn+\z)\Q=
\Qa(\Q\Qa -2\Bn -\z)^{-1}\Q=\\ \notag \Qa
\Q(\Qa\Q-2\Bn-\z)^{-1}=\PF_-\Rm(2\Bn+\z).\end{gather}

Integration gives
\begin{equation}
\nonumber
    \int_{\G_0}\Qa(\PF_- -2\Bn
    -\z)^{-1}\Q d\z=\PF_- \int_{\G_1}(\PF_-
    -\z)^{-1} d\z =2\pi i\PF_-P_1.
\end{equation}
By Theorem~\ref{2:TheoIwatsuka}, the point $\L_1=2\Bn$
is an isolated point of the essential spectrum of
$\PF_-$, therefore the nonzero eigenvalues of
$\PF_-P_1$ may converge only to $\L_1$, and thus
$(\PF_--2\Bn)P_1=\PF_-P_1-2\Bn P_1$ is compact.
Moreover, by Proposition~\ref{2:PropComp}, this
compactness is preserved after multiplication by any
product of creation and annihilation operators.

 Now we
consider the second term,  $H(\z)$ in
\eqref{4:TransRes2}.  We are going  to
 show that for any operator of the type $\Tc$
  the integral of $\Tc H(\z)\Tc$
 along  $\G_0$ is compact. First we apply the resolvent
formula \eqref{4:TransRes2} $k$ times to the first
factor in $H(\z)$, with $k$ large enough. This
operation will produce terms of order $-4$, $-6$,
\dots,
 containing  factors $\Rm(\z)$
and $\bb$, and the remainder term of order $-2k$
containing these factors with, additionally, one
factor $\Rp(2\Bn+\z)$. This last  remainder term, for
$k$ sufficiently large,
 is compact by
Lemma~\ref{4:LemmaComp1}, even after the
multiplication by $\Tc$ from the left and from the
right.

 The leading terms in $H(\z)$, having orders $-4$,
 $-6$,\dots,
will be transformed by repeatedly commuting $\bb$ and
$\Rm(\z)$ and then the resulting commutants again with
$\Rm(\z)$ and so on. Under commuting $\Rm(\z)$ with a
function, with $\Q$, or with $\Qa$, this factor
$\Rm(\z)$ moves to the left or to the right, and one
more product in the sum composing $\Tc H(\z) \Tc$
arises, of order lower by $1$. In this commuting
procedure we aim for collecting the factors $\Rm(\z)$
together. As soon as a term with all  $\Rm(\z)$
standing together arises, we leave it alone and do not
transform any more. After sufficiently many
commutations, we arrive at a collection of terms in
$\Tc H(\z)\Tc$ of negative order, which are compact by
Lemma~\ref{4:LemmaComp1}. The terms of nonnegative
order, the ones to which this Lemma cannot be applied,
  will have the form $G_1\Rm(\z)^kG_2$ with $k>1$ and
   some operators $G_1,G_2$. These terms vanish
 after integration over $\G_0$.

Now we demonstrate in  detail  how this procedure
works. Let $\Tc=1$. For the case $q=1$, which we are
considering at the moment, it suffices to apply
\eqref{4:TransRes2} twice and write
\begin{gather}\label{4:TransRes4}
    H(\z)=2\Rm(\z)\bb
\Rp(2\Bn+\z) =2\Rm(\z)\bb\Rm(\z)\\
\notag-4\Rm(\z)\bb\Rm(\z)\bb\Rp(2\Bn+\z).
\end{gather}
We multiply \eqref{4:TransRes4} by $\Qa$ and $\Q$, as
\eqref{4:ApproxProj} requires, and then the last term
takes the  form
$\Qa\Rm(\z)\bb\Rm(\z)\bb\Rp(2\Bn+\z)\Q.$ This is an
operator of order $-2$, containing as a factor  the
compactly supported function $\bb$, so it is compact
for any $\z\in \G_1$ by Lemma~\ref{4:LemmaComp1}. As
for the first term in \eqref{4:TransRes4}, we handle
it in the following way (we suppress $\z$ as long as
it does not cause a misunderstanding)
\begin{gather}\label{4:TransRes5}
\Rm\bb\Rm=\bb\Rm^2-\Rm[\PF_-,\bb]\Rm^2=\\
\notag \bb\Rm^2 -[\PF_-,\bb]\Rm^3
+\Rm[\PF_-,[\PF_-,\bb]]\Rm^3
\end{gather}
(we used here that $[\Rm, L]=-\Rm[\PF_-,L]\Rm$ for any
 operator $L$.)
Again we multiply \eqref{4:TransRes5} by $\Qa$ and $
\Q$. In the first and the second terms on the
right-hand side all entries of the resolvent are
collected together, so these terms vanish after the
integration; the last one has order $-4$ and is
compact by Lemma~\ref{4:LemmaComp1}. If the operator
$\Tc$ is present, we continue expanding
\eqref{4:TransRes4} sufficiently long:
\begin{gather}\label{4:TransRes5.1}
    H(\z)=2\Rm(\z)\bb
\Rp(2\Bn+\z) =2\Rm\bb\Rm\\
\notag-4\Rm\bb\Rm\bb\Rp(2\Bn+\z)\dots\\ \notag =
\sum_{j=1}^k (-2 \Rm\bb)^j\Rm +(-2
\Rm\bb)^{k+1}\Rp(2\Bn+\z).
\end{gather}
We choose $k$ large enough, so that the  last term in
\eqref{4:TransRes5.1} will be of negative order even
after the multiplication by $\Tc$. In all remaining
terms in \eqref{4:TransRes5.1} we  commute $\Rm$  with
other operators, moving all factors $\Rm$ say, to the
right. For example, the term $\Rm\bb\Rm\bb\Rm$ will be
handled in the following way
\begin{gather}
\nonumber
\Rm\bb\Rm\bb\Rm=\bb\Rm^2\bb\Rm+\Rm[\bb,\PF_-]\Rm^2\bb\Rm\\
\notag=
\bb\Rm\bb\Rm^2+\bb\Rm^2[\bb,\PF_-]\Rm\bb\Rm+
\Rm[\bb,\PF_-]\Rm^2\bb\Rm.
\end{gather}
The second and third terms in the last line are
operators of order $-7$, and they stay compact after
multiplication by four operators $\Q$, $\Qa$, which
gives compactness of corresponding terms in $\Tc\Qa
H(z)\Q \Tc$ for an operator $\Tc$ containing just one
factor. The first term in the last line has order
$-6$, so we have to make one more commutation:
\begin{gather}\label{4:TransRes7}
\bb\Rm\bb\Rm^2=\bb^2\Rm^3+\bb\Rm[\bb,\PF_-]\Rm^3.
\end{gather}
Now the first term in \eqref{4:TransRes7} vanishes
after integration, and the second one has order $-7$.
If the operator $\Tc$ contains more terms, we have to
continue commuting, moving the factors $\Rm$ to the
right. If there is a separately standing $\Rm$ in some
term, we can move it to the right, and the order of
the term with commutant decreases by one. We cannot do
this and, correspondingly, decrease the order, only if
all $\Rm$ stand together, but such terms vanish after
the integration. This reasoning establishes our
property for $\SF_1$.

Now we explain how one handles the operators $\SF_q$,
$q>1$.
 It is convenient to use the induction in $q$.
 So, supposing that for all $q<p$ the
compactness is proved, we start transforming the
 expression \eqref{4:ApproxProj} for $q=p$.
  First we extract the principal term. We write, using
\eqref{4:TransRes2},
\begin{equation}\label{4:TransRes8}
    \Qa^p \Rm(\z) \Q^p =\Qa^{p}(\Rp(2\Bn+\z)-H(\z))
    \Q^p.
\end{equation}
The term $H(\z)$ produces a compact contribution,
in the same way as for $p=1$.
 For the
leading term  in \eqref{4:TransRes8} we use
 \eqref{4:TransRes2.1}:
\begin{equation}
\nonumber
    \Qa^p\Rp(2\Bn+\z)\Q^p=\Qa^{p-1}\PF_-\Rm(2\Bn+\z)\Q^{p-1}.
\end{equation}
We commute $\Qa^{p-1}$ and $\PF_-$ in the last
expression to get
\begin{gather}\label{4:TransRes10}
    \Qa^p\Rp(2\Bn+\z)\Q^p =
    \PF_-\Qa^{p-1}\Rm(2\Bn+\z)\Q^{p-1}+\\ \notag
    [\Qa^{p-1},\PF_-]\Rm(2\Bn+\z)\Q^{p-1}.
\end{gather}
The first term in \eqref{4:TransRes10} has the form,
similar to the one we  have started with in
\eqref{4:TransRes8}, just with $p$ replaced by $p-1$,
and the inductive assumption can be applied. The
commutator $[\Qa^{q-1},\PF_-]$ in the second term in
\eqref{4:TransRes10} is the sum of $p-1$ expressions
of the form
\begin{equation}\label{4:TransRes11}
\Qa^{j}[\Qa,\PF_-]\Qa^{q-2-j}=2\Bn\Qa^{p-1}+2\Qa^{j+1}
\bb\Qa^{p-2-j},\; j=0,\dots, p-2.
\end{equation}
This gives additionally $p-1$ terms of the form $2\Bn
\Qa^{p-1}\Rm(2\Bn+\z)\Q^{p-1}$ in our integrand. So we
have transformed $\Qa^p\Rp(2\Bn+\z)\Qa^p$ to
$(2(p-1)\Bn+\PF_-) \Qa^{p-1}\Rm(2\Bn+\z)\Q^{p-1}$ plus
remainder terms.

After $p$ such steps  we arrive at the expression
\begin{equation}
 \nonumber   \MF\Rm(2p\Bn+\z) ;
    \;\MF=\prod_{j=1}^p(2(p-j)\Bn+\PF_-),
\end{equation}
which, after the change of variables $\z'=2p\Bn+\z$
and integration along $\G_p$, gives $2\pi i\MF P_p$.
The latter operator is a compact perturbation of $2\pi
iC_p P_p$, by Proposition~\ref{2:PropUnpertSubspace},
and, as usual, this perturbation remains compact after
multiplication by $\Tc$.

The remainder terms arising in  the expressions
similar to \eqref{4:TransRes8}, are taken care of by
means of Lemma~\ref{4:LemmaComp1}, as in the case of
$q=1$: they contain at least two resolvents separated
by a function and the commutation procedure succeeds.
Another kind of remainders  are the ones obtained in
the relations similar to \eqref{4:TransRes11},
substituted into \eqref{4:TransRes10}. They contain
only one resolvent. However after the integration such
terms take the form $\Tc' P \Tc''$ where $P$ is one of
projections $P_q$, and $\Tc', \Tc''$ are products of
creation, annihilation operators and functions,
moreover at least one of these functions has compact
support. By Lemma~\ref{4:LemmaComp2}, such operators
are compact.
\end{proof}
So we have proved that the operator $\SF_q$ is close
to the projection $P_q$.
\begin{proposition}\label{4:Prop:SclosetoQ} The
operator $\SF_q$ is close to the projection $Q_q$ onto
the approximate spectral subspace
$\Gc_q$.\end{proposition}
\begin{proof} First of all, since the operator $\SF_q$
is a self-adjoint operator close to \emph{some}
projection, it must have  the essential spectrum as
any projection has, this means, consisting only of the
points $0$ and $1$. Moreover, since the range of
$\SF_q$ lies in $\Gc_q$, the operator $\SF_q$ has
$\Gc_q$ as an invariant subspace and it annuls the
orthogonal complement to $\Gc_q$.
 So we may consider $\SF_q$ in $\Gc_q$.
We are going to show first that zero is not a limit
point of nonzero eigenvalues of  $\SF_q$;  this will
imply that  zero is an isolated point of the spectrum
of $\SF_q$ and the range of $\SF_q$ is closed in
$\Gc_q$.
 To
show  this, we consider the self-adjoint  operator
 $C_q^{-1} \GF^*\GF:\Hc_0\to\Hc_0$, $\GF=\Qa^qP_0$.
By Proposition~\ref{3:basicPropositionBP} and
Lemma~\ref{3:LemComp}, this operator is bounded, and,
moreover,
\begin{equation}\label{4:ProofClosed.1}
    C_q^{-1}\GF^*\GF  =P_0(1+\KF)P_0
\end{equation}
with a compact operator $\KF$. This implies that the
point $0$ is an isolated point  of the spectrum of
$C_q^{-1}\GF^*\GF :\Hc_0\to\Hc_0$, and therefore $0$
is an isolated point of the spectrum of $\SF_q=
C_q^{-1} \GF\GF^*$. So the range of $\SF_q$ is closed.
Further on,
 it follows from \eqref{4:ProofClosed.1} that
  $\GF^*=P_0 \Q^q:\Gc_q\to \Hc_0$
has range of finite codimension in $\Hc_0$, and since
$\Qa^q$ maps $\Hc_0$ onto $\Gc_q$ (by the definition
of the latter space), the range of $\SF_q$ has finite
codimension in $\Gc_q$.
 These properties, taken together, imply that  $\SF_q$ is
 a compact perturbation of the projection $Q_q$
 acting in the range of $Q_q$, $\SF_q=Q_q(1+\KF_1)Q_q$,
 with a compact operator $\KF_1$.
 Since the range of $\SF_q$ has finite co-dimension in
 $\Gc_q$, this relation can also be written as
  $\SF_q-Q_q=\SF_q\KF_2\SF_q+\FF$
 with a compact operator $\KF_2$ and $\FF$ having finite
 rank. By Lemma~\ref{4:LemmaComp2}, the operator $\SF_q$
 stays bounded after multiplication by any collection
 of creation and annihilation operators, and this
 establishes closeness of $\SF_q$ and $Q_q$.
\end{proof}
Propositions~\ref{4:Prop:SclosetoP}  and
\ref{4:Prop:SclosetoQ}, taken together, prove
Theorem~\ref{4:ThmCloseness}.

Now we add a perturbation by a compactly supported
smooth  electric potential $V(x)$. We are going to
show that under such perturbation the spectral
subspaces 'almost' do not change. This observation
justifies the usual physical
 intuition hinting that
 in the first order of perturbation theory only
  the eigenvalues change,
  while the spectral subspaces change only in the second
  order. Note that by our construction of the
  approximate spectral subspaces we have shown that
  this physical intuition is wrong in the case of a
  magnetic perturbation.

  Note, first of all, that the perturbation of $\PF_-$ by $V$ is
  relatively compact, therefore, again, the spectrum of the
   operator $\PF_-(V)=\PF_-+V$ between Landau levels
   is discrete. We  can change  the  contours
   $\G_q$ a little, so that they do not pass through the
   eigenvalues of $\PF_-(V)$.
We denote by $\Hc^V_q$ the spectral subspaces of
$\PF_-(V)$
 corresponding to the
spectrum inside $\G_q$, by $P^V_q$ the corresponding
spectral projections, and by $\Rm^V(\z)$ the resolvent
of $\PF_-(V)$.

\begin{proposition}\label{4:PropElPert} The projections
$P^V_q$ and $Q_q$ are close in the sense used  in
Theorem~\ref{4:ThmCloseness}.\end{proposition}
\begin{proof} We will prove that the projection
$P^V_q$ is close to $P_q$, then the result will follow
from Theorem~\ref{4:ThmCloseness}.

We use the representation of projections by means of
resolvents,
\begin{equation}
\nonumber
 P^V_q =(2\pi i)^{-1}\int_{\G_q}\Rm^V(\z)d\z.
\end{equation}
The similar expression for $P_q$ and the resolvent
formula give
\begin{gather}\label{4:PropElPert.2}
    P^V_q-P_q =-(2\pi i)^{-1}\int_{\G_q}\Rm(\z)V
    \Rm^V(\z)d\z=\\ \notag
(2\pi i)^{-1}\sum_{k=1}^{2N-1}\int_{\G_q}
    (-\Rm(\z)V)^k\Rm(\z)d\z \\ \notag+(2\pi
    i)^{-1}\int_{\G_q}(-\Rm(\z)V)^N\Rm^V(\z)(-V\Rm(\z))^N
    d\z.
\end{gather}
The last term in \eqref{4:PropElPert.2} is compact and
stays compact after the multiplication by the creation
and annihilation operators, as soon as $N$ is large
enough, by Lemma~\ref{4:LemmaComp1}.  With the leading
terms in \eqref{4:PropElPert.2}, we can perform the
same procedure as when proving
Proposition~\ref{4:Prop:SclosetoP}. We  commute the
resolvent with $V$ and with the terms arising by
commutation and so on, aiming to collect the
resolvents together all the time. As a results we
arrive at a number of terms of sufficiently negative
order, thus compact before the  integration, and terms
with all resolvents collected together, and thus
vanishing after the integration.\end{proof}

\section{Spectrum of Toeplitz-type operators}\label{Toeplitz}

We move on to the study of the splitting of Landau
levels of our operators. Similar to \cite{RaiWar},
\cite{MelRoz}, \cite{FilPush}, the properties of the
eigenvalues of the perturbed operators depend on the
properties of the spectrum of certain Toeplitz-type
operators.

Usually, by Toeplitz operator one understands an
operator of the form $\TF^W=PWP$ where $P$ is the
orthogonal projection onto some subspace $\Gc$ in
$L_2$ and $W$ is the operator of multiplication by
some function. Alternatively, if we consider the
Toeplitz operator in $\Gc$ the operator can be written
as $\TF^W=PW$.

In \cite{RaiWar}, \cite{MelRoz} such operators were
considered, with $\Gc$ being one of Landau subspaces
and the electric potential $V$ acting as $W$. Later,
in \cite{FilPush}, Toeplitz type operators at the
lowest LL were considered, with some differential
operators acting as $W$. We will study this latter
kind of Toeplitz type operators, with the space
$\Hc_0$ of zero modes of the perturbed Pauli operator
acting as $\Gc$.

We recall that   the projection onto the space
$\Hc_0=\Gc_0$ of zero modes of $\PF_-$ is denoted by
$P_0$. We consider linear differential operators $\LF$
of order $l$ with bounded coefficients,  $\LF u =
\sum_{\a+\b\le l}p_{\a\b}\partial_1^\a\partial_2^\b
u$. We suppose that all coefficients $p_{\a\b}$ have
compact support in the disk $|x|\le \Rb$ and
$|p_{\a\b}|\le \Mb$. With any such operator $\Lc$ we
associate the Toeplitz-type operator in $\Hc_0$:
\begin{equation}\label{5:Above.1}
  \nonumber  \TF^\LF v = P_0 \LF^*\LF v,\; v\in \Hc_0.
\end{equation}
This operator corresponds to the quadratic form
\begin{equation}
  \nonumber  \tF^\LF[u]=\int |\LF u|^2 dx, \;u\in \Hc_0.
\end{equation}

We are going to study the spectrum of $\TF^\LF$ . We
denote by $\l_n^{\pm}=\l_n^{\pm}(\LF)$ positive,
 resp., negative eigenvalues of the operator
 $\TF^\LF$. The distribution functions $n_\pm(\l),\, \l>0,$
 are
 defined as $n_\pm(\l)=n_\pm(\l;\LF)=$ $\#\{n:\;\pm
 \l_n^{\pm}(\LF)>\l\}$.
 The singular numbers $s_n(\LF)$
of the operator $\TF^\LF$ are just the absolute values
 of $\l_n^{\pm}$ ordered non-increasingly, and their
 distribution function equals $n(\l;\LF)=n_+(\l;\LF)+n_-(\l,\LF)$.
We denote by $\X(\l)$ the function $\frac12 \frac{|\ln
\l|}{\ln|\ln \l|}$.

\begin{proposition}\label{5:EstimateAboveL}
For an operator $\LF$ with bounded compactly supported
coefficients,
\begin{equation}\label{5:EstimateAbove.0}
    \limsup_{\l\to0} n_\pm(\l;\LF)\X(\l)^{-1}\le1.
\end{equation}
\end{proposition}

The estimate \eqref{5:EstimateAbove.0} means that the
eigenvalues of $\TF^{\LF}$ converge to zero extremely
rapidly, super-exponentially:
\begin{equation}\label{5:EstimateAbove.01}
    \limsup_{n\to\infty}(n!\l_n^\pm(\LF)
    )^{\frac1{n}}<\infty.
\end{equation}
 The estimates \eqref{5:EstimateAbove.0}, \eqref{5:EstimateAbove.01}
  do not exclude the possibility
that the operator $\TF^{\LF}$ has only a finite number
of positive and/or negative eigenvalues, counting
multiplicity.
\begin{proof} We start by proving a special version of
the proposition, dealing with the case when the
operator $\LF$ is just the multiplication by a bounded
compactly supported  function $W$. The corresponding
Toeplitz operator will be denoted by $\TF^W$

 We apply the classical variational approach.
 Consider the quadratic form $\tF^W[v]$ of
the operator $\TF^W$:
\begin{equation}
\nonumber
    \tF^W[v]=\int W(x)|v(x)|^2 dx,\; v\in \Hc_0.
\end{equation}
The variational description of eigenvalues (Glazman
lemma) implies
\begin{equation}
\nonumber
 n_+(\l;W)= \min\codim\{\Ls\subset \Hc_0: \tF^W[v]\le
\l \|v\|^2, \, v\in\Ls\}.
\end{equation}
We recall the description of the space $\Hc_0$ to get
\begin{equation}\label{5:EstimateAbove.3}
n_+(\l;W)=
\min\codim\{\Ms\subset\Fc^\bb:\tF^W[e^{-\P}f]\le \l
\|e^{-\P}f\|^2, f\in\Ms\},
\end{equation}
where $\Fc^\bb$ is the space of entire analytical
functions, defined in Sect.2.

Now we are going to estimate the quantity in
 \eqref{5:EstimateAbove.3}
from above by means of changing the forms and the
spaces. First, if we replace $\|e^{-\P}f\|^2$ in
 \eqref{5:EstimateAbove.3} by
something smaller, then there will be fewer subspaces
$\Ms$ where the inequality holds and therefore $\min
\codim$ may become only larger. So we can replace
$\|e^{-\P}f\|^2$ by $C\|e^{-(1+\e)\P^\circ}f\|^2$
where $\P^\circ$ is the potential defined in
\eqref{2:Creation.00} and $C$ is some constant, since,
 due to our
 estimates of $\Psi$, discussed  in Sect.3,
  $e^{-\P}\ge C e^{-(1+\e)\P^\circ}$. After
this, if we replace the enveloping space $\Fc^\bb$  in
\eqref{5:EstimateAbove.3} by a larger one,
$\Fc_{(1+\e)\Bn}$, then the codimension of subspaces
may only grow. Therefore, we arrive at
\begin{gather}\label{5:EstimateAbove.4}
n_+(\l,W)\le\\ \notag
\min\codim\{\Ms\subset\Fc_{(1+\e)\Bn}:\tF^W[e^{-\P}f]\le
\l \|C e^{-(1+\e)\P^\circ}f\|^2, f\in\Ms\}.
\end{gather}
Now we notice that $\tF^W[e^{-\P}f]=\int
We^{-2\P}|f|^2 dx$, and  $We^{-2\P}$ is again a
bounded function with compact support. Therefore we
can apply estimates for the right-hand side in
\eqref{5:EstimateAbove.4} established in
\cite{RaiWar}, Proposition~3.2 or \cite{MelRoz},
Lemma~6.1, which lead to the inequality
\eqref{5:EstimateAbove.0}.

 The  estimate we have just established
can even be improved, if we use the results of
\cite{FilPush} to majorize  the right-hand side in
\eqref{5:EstimateAbove.4}:
\begin{equation}
\nonumber
 \limsup_{n\to\infty}(n!\l_n^\pm(W)
    )^{\frac1{n}}\le \frac{\Bn}2 \Cp(\supp W),
\end{equation}
where $\Cp(\supp W)$ is the logarithmical capacity of
the support  of $W$.

Now we pass to the general case. Let all coefficients
$p_{\a\b}$ of the operator $\LF$ have support in the
disk with radius $\Rb$ centered at $0$ and
$|p_{\a\b}|\le \Mb$.

Denote by $\chi$ the characteristic
 function of the disk $\Db$ of radius $2\Rb$.
By Cauchy integral formula, applied as in
Lemma~\ref{3:LemComp}, each term of the form
$p_{\a\b}\partial_1^{\a}\partial_2^{\b}u$, $u\in
\Hc_0$ can be expressed as $T_{\a\b}( \chi u)$, with a
bounded operator $T_{\a\b}$. The upper bound for the
norm of this operator depends only on the order of
derivatives $\a+\b$, on the size of the coefficient
$p_{\a\b}$ and on the radius $\Rb$. Since the
quadratic form $\tF^\LF$ of the operator $\TF^\LF$ has
the form $\tF^\LF[u]=\|\sum
p_{\a\b}\partial_1^{\a}\partial_1^{\b}u\|^2$ $=\|\sum
p_{\a\b}\partial_1^{\a}\partial_1^{\b}\chi u\|^2$, we
have
\begin{equation}\label{5:Above.4}
|\tF^\LF[u]|\le \varkappa^{-1} \|\chi
u\|^2=\varkappa^{-1}\tF^\chi[u], \; u\in \Hc_0
\end{equation}
for a certain $\varkappa$. Thus the singular numbers
of the operator $\TF^\LF$ are majorated by the ones of
the operator $\TF^W$ with  $W=\chi,$ and for such
operators the required estimate is already proved.
\end{proof}

It is more convenient for our needs  to formulate
Proposition~\ref{5:EstimateAboveL},  directly in the
variational form, declaring the existence of subspaces
with prescribed properties.

\begin{proposition}\label{5:PropAbove}
There exists  a  constant $\varkappa$ depending only
on $l, \Rb, \Mb$, such that for any $\e>0$ and for any
$\l>0,$ sufficiently small, there exists a subspace
$\Ms(\l)\in \Hc^0$ such that the codimension of
$\Ms(\l)$ is no greater than $(1+\e)\X(\l)$, and for
any order $l$ differential operator $\LF$ with
coefficients $p_{\a\b}$ supported in the disk with
radius $\Rb$ and satisfying $|p_{\a\b}|\le \Mb$
\begin{equation}\label{5:Above.3}
 \tF^\LF[u]\le \varkappa\l\|u\|^2, u\in \Ms(\l).
\end{equation}
\end{proposition}

  The important  feature in this formulation is
that the subspace $\Ms(\l)$ of controlled codimension
can be chosen in such way that it services
simultaneously all Toeplitz-like operators generated
by differential operators $\LF$ of fixed order and
with coefficients subject to the above restrictions.

\begin{proof} By \eqref{5:Above.4}, as soon as the
chosen subspace $\Ms(\l)$ services the Toeplitz
operator $\TF^\chi$, the inequality \eqref{5:Above.3}
holds for all operators $\LF$ subject to our
conditions.\end{proof}
 We formulate the eigenvalue
estimate from below for operators $\TF^W$ only, under
the condition that  the function $W$ is non-negative.
This latter condition can be somewhat relaxed, however
some kind of positivity requirements still remain. In
fact, starting from \cite{RaiWar}, \cite{MelRoz}, the
positivity condition for  lower spectral estimates
remains a serious obstacle in this field.

\begin{proposition}\label{5:EstimateBelow}
Let the function $W$ be non-negative and  be greater
than some $c>0$ on an open set. Then
\begin{equation}
  \nonumber  \liminf_{\l\to0} n_+(\l;W)\X(\l)^{-1}\ge 1.
\end{equation}
\end{proposition}
\begin{proof}
We use the variational description of $ n_\pm(\l;W)$:
\begin{equation}\label{5:EstimateBelow.1}
 n_+(\l;W)= \max\dim\{\Ls\subset \Hc_0: \tF_0[v]<
\l \|v\|^2, \, v\in\Ls, v\ne0\}.
\end{equation}
Again, using the description of the space $\Hc_0$ of
zero modes,
 we can rewrite \eqref{5:EstimateBelow.1} as
\begin{equation}\label{5:EstimateBelow.2}
n_+(\l;W)=
\max\dim\{\Ms\subset\Fc^\bb:\tF_0[e^{-\P}f]> \l
\|e^{-\P}f\|^2, f\in\Ms, f\ne 0\}
\end{equation}
By replacing $\|e^{-\P}f\|^2$ by some larger quantity,
we narrow the choice of subspaces $\Ms$ where the
inequality in \eqref{5:EstimateBelow.2} holds. So, we
replace $\|e^{-\P}f\|^2$ by
$\|Ce^{-(1-\e)\P^\circ}f\|^2$ with an arbitrary
$\e\in(0,1)$ and some $C$. This leads to decreasing of
the right-hand side of
 \eqref{5:EstimateBelow.2}. A further decreasing is
 obtained by narrowing the enveloping space: we
 replace $\Fc^\bb$ by $\Fc_{(1-\e)\Bn}$. This
 transforms the right-hand side in \eqref{5:EstimateBelow.2}
 to the form, where the estimate from below is already
 proved, again in  \cite{RaiWar}, Proposition~3.2 or
\cite{MelRoz}, Lemma~6.1.
\end{proof}

We will need the uniform variational version of the
latter estimate.

\begin{proposition}\label{5:ToeplitzSubspaces} Let $\Omega$
 be an open set, $c>0$.
Then for any $\e>0$ and
  for any $\l>0$, small enough, depending on $\e$
there exists a subspace $\Ls=\Ls(\l)\subset \Hc_0$
such that $(Wv,v)> \l \|v\|^2, \,
v\in\Ls\setminus\{0\}$ for all functions $W\ge0$
satisfying $W\ge c$ on $\Omega$ and  $ \dim \Ls(\l)\ge
(1-\e)\X(\l)$.\end{proposition}

The result follows from the estimate $(Wv,v)\ge
c(\chi_\Omega v,v)$.

\section{Perturbed eigenvalues}

 The passage
from the spectral estimates for Toeplitz-type
operators to the ones for the Pauli operator has been
performed in a different way
 in \cite{Raikov1}, \cite{RaiWar}  and in  \cite{MelRoz}.
A short and neat reasoning in \cite{Raikov1},
\cite{RaiWar} based upon the proper version of the
Birman-Schwinger principle requires positivity of the
perturbing operator  which fails to be positive in our
case. Therefore we  follow the approach of
\cite{MelRoz}, Sect.~9, using variational
considerations, however certain new features appear.

The particular form of the variational principle we
use, derived immediately from the spectral theorem, is
the following. For a self-adjoint operator $\AF$ and
an interval $(s,r)$ in the real axis we denote by
$N(s,r ; \AF)$ the total multiplicity of the spectrum
of $\AF$ in $(s,r)$; when this quantity is finite, it
is the number of eigenvalues of $\AF$ in $(s,r)$,
counting multiplicities.

\begin{proposition}\label{6:VarPrinc}
Let $\AF$ be a self-adjoint operator with domain
$\dom{\AF}$. Then, for $s < r$,
\begin{eqnarray}
\lefteqn{ \hspace*{-0.7cm} N(s,r ; \AF) } \nonumber \\
& & \hspace*{-1.55cm} = \max  {\rm dim}  \left\{ {\Ls}
 \subset  \dom{\AF} :  \| (\AF-\m)u \|^{2}  <  \t^{2}
 \| u \|^{2},  u \in  {\Ls} \backslash \{ 0 \} \right\} \label{6:VarPrinc.01} \\
& &  \hspace*{-1.55cm} = \min {\rm codim} \! \left\{
{\Ls} \! \subset \! \dom{\AF} : \| (\AF\!-\!\m)u\|^{2}
\!
 \geq \!\t^{2} \| u \|^{2}, u \! \in \! {\Ls} \right\},
\label{6:VarPrinc.02}
\end{eqnarray}
where $\m=(r+s)/2$ and $\t=(r-s)/2$.
\end{proposition}

When using Proposition~\ref{6:VarPrinc}, one looks for
a subspace $\Ls$ where the inequality in
\eqref{6:VarPrinc.01} is fulfilled.  The dimension of
this subspace gives an estimate from below for $N(s,r
; \AF)$. On the other hand, having found a subspace
$\Ls$ where the inequality in  \eqref{6:VarPrinc.02}
is satisfied, we can be sure that the codimension of
this subspace estimates $N(s,r ; \AF)$ from above.

Considering   perturbations of the Landau Hamiltonian
by general electric and magnetic fields, we should
accept the possibility that the eigenvalues may split
away from the Landau level both up and down. Therefore
we need rather advanced notations. We fix
 a Landau level $\L=\L_q=2q\Bn$ and set $s_\pm=\L\pm \Bn$.
For given $\l\in(0,\Bn/4)$ we define $\m_\pm=(\L\pm\l
+s_\pm)/2$, $\t= \frac12\Bn-\frac{\l}{2}$. Thus,
$\m_\pm$ is the midpoint of the interval between
$\L\pm\l$ and $s_\pm$, $\t$ is the half-length of this
interval. For the perturbation given by an operator
$\VF$ we denote by $N_+(\l;q,\bb,\VF)$ the eigenvalue
counting function $N(\L_q+\l,s_+;\PF_-+\VF)$ where
$\PF_-$ is the Pauli operator with magnetic field
$\Bb=\Bn+\bb$; similarly, $N_-(\l;q,\bb,\VF)$ is
defined as $N(s_-,\L_q-\l;\PF_-+\VF)$. Thus, $N_+$,
$N_-$ count the eigenvalues converging to $\L$ from
above, resp., from below.

\subsection{Upper estimate}\label{Above}
We start with the upper estimate. It declares that the
eigenvalues split away from the Landau level at least
superexponentially rapidly. This estimate is rather
robust: is does not require much smoothness of the
perturbing magnetic field or of the electric
potential. It is convenient to keep all the previous
notations in the paper for a smooth case, and use the
superscript $\sharp$ for the non-smooth versions.

 So, let $\bs\in C^2_0(\R^2)$ be a magnetic field with
compact support, $\ps\in C^4$ be its scalar potential,
$\D\ps=\bs$, and $\as\in C^3$ its vector potential.
The Pauli
 operator associated to the field $\Bn+\bs$ will be denoted by
 $\PF_-^\sharp$.

 We consider the
smoothened magnetic field, $
    \bb=\omega*\bs $,
where  $\omega\in C_0^\infty$ is a smooth function
with compact support, $\int \omega dx=1$, and
$\omega(x)$ depends only on $|x|$. Thus the field
$\bb$ is smooth and compactly supported. Of course,
$\p=\omega*\ps$ is a potential for $\bb$, $\D\p=\bb$.
Moreover, since $\ps$ is a harmonic function outside
the support of $\bs$, by the mean value property, $\p$
coincides with $\ps$ outside some compact.
Correspondingly the vector potential $\ab$ of the
field $\bb$ coincides with $\as$ outside a compact.

As a result, the Pauli operator $\PF_-^\sharp$ can be
expressed as
\begin{equation}\label{6:Upper.Pert1}\nonumber
\PF_-^\sharp=\PF_-+\MF,
\end{equation}
where $\PF_-$ is the Pauli operator with smooth
magnetic field $\Bb=\Bn+\bb$ and $\MF$ is a first
order differential operator with $C^2$-coefficients
having compact support.

Further on,  we consider the operator $\PF_-^\sharp+V$
as a perturbation of $\PF_-$,
\begin{equation}\label{6:Upper.Pert2}\nonumber
\PF_-^\sharp+V=\PF_-+\VF,\; \VF=\MF+V, \; V\in
C_0^2(\R^2).
\end{equation}

Now we can formulate the main upper estimate.

\begin{theorem}\label{6:Upper} If the perturbations  $\bs$ and $V$ belong to $C^2_0(\R^2)$  then
\begin{equation}\label{6:Upper.0}
    \limsup_{\l\to
    0+}N_\pm(\l,\PF_-^\sharp+V)\X(\l)^{-1}\le1.
\end{equation}
\end{theorem}
\begin{proof} We will prove the 'minus' version of the
Theorem; the 'plus' version differs just by
non-essential  details. Keeping in mind
Proposition~\ref{6:VarPrinc}, we are going to
construct a subspace $\Ls$ in the domain of the
operator $\PF_-$ such that the inequality
\begin{equation}\label{6:Upper.1}\nonumber
    \|(\PF_-+\VF-\m)u\|^2
\ge \t^2\|u\|^2\end{equation} holds for all $u\in\Ls$,
and estimate its codimension.

We represent an arbitrary function $u$ as $u=u_1+u_2$
where $u_1\in \Gc=\Gc_q$ belongs to the approximate
spectral subspace constructed in
Section~\ref{Subspaces3} and $u_2$ belongs to the
subspace $\hat{\Gc} $ orthogonal to $\Gc$, $u_1=Q_qu$,
$u_2=\hat{Q}_qu$, $\hat{Q}_q=1-Q_q$.

Then the inequality we aim for takes the form
\begin{equation}\label{6:Upper.2}
\|(\PF_-+\VF-\m)(u_1+u_2)\|^2-\t^2\|u_1\|^2-\t^2\|u_2\|^2\ge
0.
\end{equation}
The left-hand side in \eqref{6:Upper.2}
can be written as
\begin{gather}\notag
 \bl\|(\PF_-+\VF-\m)u_1\|^2-\t^2\|u_1\|^2\br +
\bl\|(\PF_-+\VF-\m)u_2\|^2-\t^2\|u_2\|^2\br\\
\label{6:Upper.3}
+2\Re((\PF_-+\VF-\m)u_1,(\PF_-+\VF-\m)u_2).
\end{gather}
Following the pattern of the proof of Proposition 9.2
in \cite{MelRoz}, we are going to  find subspaces of
controlled codimension in $\Gc$ and in $\hat{\Gc}$,
where the first and the second terms in
\eqref{6:Upper.3} are positive, with some margin,
while the third term in \eqref{6:Upper.3} is majorized
by this margin.

\textbf{The first term.} We start with the first term.
 Since $u_1\in \Gc$, by
our construction, $u_1$ can be written in a unique way
as $u_1=\Qa^q v_1$, $v_1\in \Hc_0$, and thus the term
takes the form
\begin{gather}\label{6:Upper.4}
    \|(\PF_-+\VF-\m)u_1\|^2-\t^2\|u_1\|^2=\|(\PF_-+\VF-\m)\Qa^q v_1\|^2
    -\t^2\|\Qa^q
    v_1\|^2\\ \notag
=\|(\PF_--\m)\Qa^qv_1\|^2-\t^2\|\Qa^q
    v_1\|^2 +2\Re((\PF_--\m)\Qa^q v_1, \VF\Qa^q v_1)\\
    \notag+(\VF\Qa^q v_1,\VF\Qa^q
    v_1).
\end{gather}
 The first two
terms in \eqref{6:Upper.4}  can be transformed by
means of Proposition~\ref{3:basicPropositionBP}; they
produce $C_q((\L-\m)^2 -\t^2)\|v_1\|^2$ as well as
$C_q(W_1v_1,v_1)$ with some compactly supported
function $W_1$.
 The last term is non-negative and can be ignored when
  estimating the whole expression from below. As
 for the remaining, next to last term
 in \eqref{6:Upper.4}, we have
\begin{equation}\label{6:Upper.4.1}\nonumber
2\Re((\PF_--\m)\Qa^q v_1, \VF\Qa^q v_1)\ge
-(\|\chi\LF_1v_1 \|^2 +\|\LF_2v_1 \|^2),
\end{equation}
where $\chi\LF_1,\LF_2$ are order $q+2$ and order
$q+1$ differential operators with bounded compactly
supported coefficients, $\LF_1=(\PF_--\m)\Qa^q$,
$\LF_2=\VF\Qa^q$, and $\chi$ is a characteristic
function of an open set containing the support of
coefficients of $\VF$, so that $\chi\VF=\VF$.

 As a result, we obtain \begin{gather}\label{6:Upper.5}
\|(\PF_-+\VF-\m)u_1\|^2-\t^2\|u_1\|^2\ge C_q((\L-\m)^2
-\t^2)\|v_1\|^2\\ \nonumber+(Wv_1,v_1)-\|\chi\LF_1v_1
\|^2 -\|\LF_2v_1 \|^2.
\end{gather}
 The expression $((\L-\m)^2
-\t^2)$ equals $(\L-\m-\t)(\L-\m+\t)=\l (\L-s)$. Now
we are going to chose our first subspace. We fix
$\e\in(0,\frac13)$.  By Proposition~\ref{5:PropAbove},
 there exists a subspace $\Ms_1$ in $\Hc_0$ such that
$|(Wv_1,v_1)|, \|\chi\LF_1v_1 \|^2, \|\LF_2v_1 \|^2$
are all no greater than $\frac1{10}\l (\L-s)\|v_1\|^2$
and the codimension of $\Ms_1$ in $\Hc_0$ is less than
$(1+\e)\X(\frac1{10}(\L-s)\l)$ for $\l$ small enough.
So, for $v_1\in \Ms_1$ the right-hand side of
\eqref{6:Upper.5} is no less than
$\frac34C_q(\L-s)\l\|v_1\|^2$. This means that on the
subspace $\Ls_1=\Qa^q\Ms_1$, which has the same
codimension in $\Gc$ as $\Ms_1$ has in $\Hc_0$, the
left hand side of \eqref{6:Upper.5} is no less than
$\frac34C_q(\L-s)\l\|v_1\|^2$. The operator
$\Qa^q:\Hc_0\to\Gc$ is boundedly invertible, so we can
majorize $\|v_1\|$ by $||u_1||$, as in
Corollary~\ref{3:Corollary}, therefore, for some
constant $c_1$, not depending on $\e,\l$,
\begin{equation}
\nonumber
 \|(\PF_-+\VF-\m)u_1\|^2-\t^2\|u_1\|\ge c_1
\l\|u_1\|^2, \; u_1\in\Ls_1.
\end{equation}
The subspace  $\Ls_1$ has codimension no greater  than
$(1+\e)\X(\frac1{10}(\L-s)\l)$ in $\Gc$.

\textbf{The second term}. In the next term in
\eqref{6:Upper.3}, $\|(\PF_--\m+\VF)u_2\|^2$
$-\t^2\|u_2\|^2$, we can write
$u_2=(1-Q_q)u_2=(1-P_q)u_2+(P_q-Q_q)u_2$ where $P_q$
is the spectral projection of $\PF_-$ corresponding to
the neighborhood $\d_q=(\L-\g,\L+\g)$ of $\L=\L_q$,
defined in Section 3. We denote by $\tilde{P}$ the
spectral projection of $\PF_-$ corresponding to the
set $(s-\Bn/2,\L-\g]\cup[\L+\g,\L+\Bn)$. Since the
spectrum of  $\PF_-$ is discrete in $(\L-2\Bn,
\L)\cap(\L,\L+2\Bn)$, the projection $\tilde{P}$ has
finite rank, and thus $u_2$ can be expressed as
\begin{equation}\label{6:Upper2.1}
u_2=(1-P_q-\tilde{P})u_2+(P_q-Q_q+\tilde{P})u_2 =g+h.
\end{equation}
Constructed as above,
 the function $g$ is lying in the spectral subspace of $\PF_-$
  corresponding to the exterior of the interval
  $(\L-3\Bn/2,\L+\Bn)$. So, by means of the spectral theorem, we can estimate
  $\|(\PF_--\mu)g\|$ from below as
      $\|(\PF_--\mu)g\|\ge
     (\t+\Bn/2)\|g\|$.
 Therefore
\begin{equation}\label{6:Upper2.2}
\|(\PF_--\mu)g\|^2-\t^2\|g\|^2\ge c_2\|g\|^2,
\end{equation}
as well as
\begin{equation}\label{6:Upper2.3}
\|(\PF_--\mu)g\|^2-\t^2\|g\|^2\ge
c_3\|(\PF_--\mu)g\|^2
\end{equation}
with some constants $c_2,c_3$ not depending on $\l$.

Now let us consider  $h=(P_q-Q_q+\tilde{P})u_2$. As we
have proved in Section~\ref{Projections4}, the
operator $P_q-Q_q$ is compact and remains compact
after multiplication by any combination of creation
and annihilation operators. This property remains
intact after adding the finite rank operator
$\tilde{P}$. So, we can write \eqref{6:Upper2.1} as
\begin{equation}\label{6:Upper2.5}
    u_2=g+\KF u_2
\end{equation}
with a compact operator $\KF$. Therefore, on  a
subspace of finite
 codimension,
\eqref{6:Upper2.5} can be rewritten as
\begin{equation}
\nonumber
    u_2=g+\KF'g
\end{equation}
with a compact operator $\KF'$ which, again, remains
compact after multiplication by $\Q,\Qa$. As a result,
for $\e$ fixed, we can find   a subspace of finite
codimension in $\Gc'\cap \Dom(\PF_-)$ such that
\begin{equation}
  \nonumber
   \frac{ \|u_2\|}{\|g\|}\in(1-\e,1+\e),\;
   \frac{\|(\PF_--\m)u_2\|}{\|(\PF_--\m)g\|}\in(1-\e,1+\e).
\end{equation}
Therefore the inequalities of the form
\eqref{6:Upper2.2}, \eqref{6:Upper2.3} can be written
for $u_2$ as well, just with a slight worsening of
constants, the value of which is not important:
\begin{equation}
\nonumber
 \|(\PF_--\mu)u_2\|^2-\t^2\|u_2\|^2\ge
c_4\|u_2\|^2,
\end{equation}
as well as
\begin{equation}\label{6:Upper2.9}
\|(\PF_--\mu)u_2\|^2-\t^2\|u_2\|^2\ge
c_5\|(\PF_--\mu)u_2\|^2.
\end{equation}

Now we can consider the complete second term in
\eqref{6:Upper.3}. This expression can be written as
\begin{gather}\label{6:Upper2.10}
    \|(\PF_-+\VF-\m)u_2\|^2-\t^2\|u_2\|^2=
     \|(\PF_--\m)u_2\|^2-\t^2\|u_2\|^2\\ \notag
    +\|\VF u_2\|^2+2\Re((\PF_--\m)u_2,\VF u_2), \; u_2\in\hat{\Gc}.
\end{gather}
The form $\|\VF u_2\|^2$ is compact with respect to
$\|(\PF_--\m)u_2\|^2$, therefore on a subspace of
finite codimension we have $\|\VF u_2\|^2<\frac{c_5}4
\|(\PF_--\m)u_2\|^2$. Further on, by Cauchy
inequality,\begin{equation}\label{6:Upper2.11}
|2\Re((\PF_--\m)u_2,\VF u_2)|\le
\frac{c_5}{4}\|(\PF_--\m)u_2\|^2 +\frac{4}{c_5}\|\VF
u_2\|^2.\end{equation} Again, using the same relative
compactness, we can restrict ourselves to a subspace
of finite codimension
 so that the last term in \eqref{6:Upper2.11} is no
greater than $\frac{c_5}4\|(\PF_--\mu)u_2\|^2$. After
substituting these estimates into \eqref{6:Upper2.10}
and recalling \eqref{6:Upper2.9}, we obtain
\begin{equation}\label{6:Upper2.12}
 \|(\PF_-+\VF-\m)u_2\|^2-\t^2\|u_2\|^2\ge
 \frac{3c_5}{4}\|(\PF_--\mu)u_2\|^2\ge c_6 \|u_2\|^2
\end{equation}
on a subspace $\Ls_2$ of finite codimension in
$\Gc'\cap\Dom(\PF_-)$. Note that the subspaces here do
not depend on the value of $\l$.

\textbf{The third term.} We are going to show now that
the third term in \eqref{6:Upper.3} is majorized by
the sum of two first  ones. We can write it in the
form
\begin{equation}\label{6:Upper3.1}
   2\Re ((\PF_-+\VF-\m)u_1,(\PF_-+\VF-\m)u_2)=
   2\Re((\PF_-+\VF-\m)^2u_1,u_2).
\end{equation}
The function $u_1\in\Gc_q$ can be written as
$u_1=\Qa^q v_1$ for some $v_1\in \Hc_0$. So we have
\begin{gather}\label{6:Upper3.2}
(\PF_-+\VF-\m)^2u_1=(\Qa\Q -\mu)^2 \Qa^q v_1
+(\PF_--\mu)\VF u_1+\\ \notag \VF(\PF_--\mu) u_1 +V^2
u_1 = (\Qa\Q -\mu)^2 \Qa^q v_1 +G.
\end{gather}
We start with the first term in \eqref{6:Upper3.2}.

In the expression $\Qa\Q\Qa^q v_1$ we perform our
usual commuting procedure moving $\Q$ to the right all
the time. As soon as $\Q$ reaches the utmost right
position, the corresponding term vanishes. Thus we
arrive at the representation
\begin{gather}\label{6:Upper3.3}
\Qa\Q\Qa^q v=2\sum\Qa^j(\Bn+\bb)\Qa^{q-j}v_1\\
\notag=2q\Bn \Qa^qv_1
+\sum_{j=1}^q\Qa^j\bb\Qa^{q-j}v=\L u_1 + F_1.
\end{gather}
In a similar way, we  transform
  the expression $\Qa\Q\Qa\Q u_1= \Qa\Q\Qa\Q \Qa^q
 v_1$: we move both copies of the
 operator $\Q$ to the utmost right position, where
 they vanish, being applied to $v_1\in \Hc_0$. What is
 left has the form
 \begin{equation}\label{6:Upper3.4}
\Qa\Q\Qa\Q \Qa^q
 v_1 =\L^2 u_1 +F_2.
\end{equation}
The functions $F_1,F_2$ in \eqref{6:Upper3.3},
\eqref{6:Upper3.4} are compositions of several, no
more than $q$, operators $\Qa$ and at least one
function with compact support, applied to $v_1$.  The
term $G=((\PF_--\mu)\VF + \VF(\PF_--\mu)  +\VF^2)
\Qa^q v_1$
 in \eqref{6:Upper3.2} involves the differential operator of order $q+2$
 with bounded compactly supported coefficients, applied to $v_1$. So we have
\begin{equation}
\nonumber
 (\PF_-+V-\m)^2u_1 =(\L-\m)^2u_1-2\m F_1+F_2+G.
\end{equation}
We substitute this expression into \eqref{6:Upper3.1}
and use the fact that $u_1$ and $u_2$ are orthogonal.
This leads us to
\begin{gather}\label{6:Upper3.6}\nonumber
2|\Re((\PF_-+V-\m)^2u_1,u_2)|=2|\Re((-2\m
F_1+F_2+G),u_2)|\\ \notag \le M\|-2\m
F_1+F_2+G\|^2+M^{-1}\|u_2\|^2,
\end{gather}
where $M>0$ can be chosen arbitrarily. We fix $M$
larger than $16 c_5^{-1}$, where $c_5$ is the constant
in \eqref{6:Upper2.12}. After this, we take $\e$
smaller than $\frac{c_5}{16M}$. By
Proposition~\ref{5:PropAbove} we can choose the
subspace $\Ms_1\subset\Hc_0$, constructed when
considering the first term in \eqref{6:Upper.2}, in
such way that the codimension of $\Ms_1$ in $\Hc_0$ is
  no greater than $(1+\e)\X(\varkappa\l)$ for some $\kappa$ and
that $\|-2\m F_1+F_2+G\|^2$ is no greater than $
\frac{c_1}{2}\l\|v_1\|^2$. With all parameters chosen
in this way, for $u_1\in\Ls_1=\Qa^q\Ms_1$ and
$u_2\in\Ls_2$ the third term in \eqref{6:Upper.3} is
majorated by the sum of the first two terms, and thus
the whole expression in \eqref{6:Upper.3} is
non-negative for $u=u_1+u_2\in \Ls_1\oplus\Ls_2$. The
subspace $\Ls=\Ls_1\oplus\Ls_2$ has codimension no
greater than $(1+\e)\X(\kappa\l)+N(\e)$. As a result,
due to $\e$ being arbitrary and to the fact that
$\X(\kappa\l)\equiv \X(\l), \l\to0$, we have
\begin{equation}
\nonumber
    \limsup_{\l\to0}N_-(\l;q,\bb,\VF)\X(\l)^{-1}=
    \limsup_{\l\to0}N_-(\l;q,\bb,\VF)\X(\varkappa\l)^{-1}\le(1+\e)
\end{equation}
Finally, we let $\e\to0$, and obtain
\eqref{6:Upper.0}.
\end{proof}
\subsection{Lower estimate}\label{below}
 Now we establish an estimate for $N_\pm(\l)$ from below. The
 conditions for this estimate to hold are expressed in
 the terms of an effective potential constructed for
 each LL from the magnetic field $\bb$ and the  electric
 potential $V$. The
 complicated form of this effective potential reflects
 the complicated character of the influence of the
 magnetic perturbation onto the behavior of
 eigenvalues: we  remind again that the latter
 dependence is not monotone.  Our construction
 requires considerable smoothness of $\bb$ and $V$;
  in order to avoid extra technicalities, we
 suppose here that both are infinitely smooth.

 For a fixed LL $\L=\L_q$, with the same  $s_\pm,\m_\pm,\t$ as
 before, we
 define the effective potentials as
\begin{gather}\label{6:Lower.00}
   \!\!\!  W_\pm=W_\pm[\bb,V,\l]=\!-\!(\L\pm\l\!+\!2\Bn)(s+2\Bn)\Zc_q[\bb]
      -\Zc_{q+2}[\bb]\\\nonumber+2(\m_\pm+3\Bn)\Zc_{q+1}[\bb]
    -\Xc_q[\bb,
    4(2\Bb-\bb+\m_\pm)\bb-(4\Bb-2\m_\pm+V)V]\\\nonumber-2\Xc_{q+1}[\bb,V-3\bb]-4\Im\Yc_q[\bb,\pd V-2\pd
    b],
\end{gather}
where the expressions $\Zc, \Xc, \Yc$ are described in
Proposition~\ref{3:basicPropositionBP}.

 \begin{theorem}\label{6:Lower}  Suppose that
for $\l$ close to zero the potential $W_\pm$ is
non-negative and, moreover,  there exists an open set
$\Omega$ where $W_\pm\ge c>0$ for $c$ independent on
$\l$.
 Then
\begin{equation}\label{6:Lower.0}
    \liminf_{\l\to0}N_{\pm}(\l,q,\bb,V)\X(\l)^{-1}\ge 1
\end{equation}
\end{theorem}
\begin{proof} We consider the eigenvalues below $\L$
and prove the estimate \eqref{6:Lower.0}
with 'minus' sign. The other case is proved in an almost
identical way.

 We are going to construct the subspace
$\Ls$ in $\Gc=\Gc_q$, having dimension asymptotically
greater than  $(1-\e)\X(\l)$ such that the inequality
\begin{equation}\label{6:Lower.1}
 \|(\PF_-+V-\m)u\|^2
< \t^2\|u\|^2
\end{equation}
is satisfied for all $u\in \Ls\setminus \{0\}.$ To do
this, we set $u=\Qa^q v$, $v\in \Hc_0$, and thus
reduce both parts of \eqref{6:Lower.1} to quadratic
forms on the lowest Landau level  using
Proposition~\ref{3:basicPropositionBP} and then apply
Proposition~\ref{5:EstimateBelow}.

So, we have
\begin{gather}\label{6:Lower.2}
\|(\PF_-+V-\m)u\|^2 -
\t^2\|u\|^2=(\PF_-u,\PF_-u)+2\Re(Vu,\PF_-u)+\\
\nonumber
(\m^2-\t^2)\|u\|^2+(Vu,Vu)-2\m(Vu,u)-2\mu(\PF_-u,u).
\end{gather}
We substitute $u=\Qa^q v,\; v\in \Hc_0$  into
\eqref{6:Lower.2} and consider each term. First, by
\eqref{3:BP:equation}, we have
\begin{equation}\label{6:Lower.3}
(\m^2-\t^2)\|u\|^2=(\m^2-\t^2)C_q\|v\|^2+(\m^2-\t^2)(\Zc_q[\bb]v,v).
\end{equation}
By \eqref{3:BP:equation.V} applied for $U=V$ and then
for $U=V^2$, we obtain
\begin{equation}\label{6:Lower.4}
-2\m(Vu,u)=-2\m(\Xc_q[\bb,V]v,v);\;
(Vu,Vu)=(\Xc_q[\bb,V^2]v,v).
\end{equation}
For the last term in \eqref{6:Lower.2}, we have
\begin{gather}\label{6:Lower.5}
-2\mu(\PF_-u,u)=-2\mu(\Qa\Q u,u)=-2\mu(\Q\Qa
u,u)+4\mu\Bn(u,u)\\ \nonumber +4\mu(\bb u,u)= -2\mu
C_{q+1}\|v\|^2\\ \nonumber-2\mu
(\Zc_{q+1}[\bb]v,v)+4\mu\Bn
C_q\|v\|^2+4\mu\Bn(\Zc_{q}[\bb]v,v)+4\m
(\Xc_q[\bb,\bb]v,v).
\end{gather}
Further on, commuting $\Q,\Qa$, we get
\begin{gather}\label{6:Lower.6}
2\Re(Vu,\PF_-u)=2\Re(Vu, \Qa\Q u)=2\Re(\Qa Vu,\Qa u
)-4(\Bb V u,u)\\ \nonumber =2( V\Qa u,\Qa
u)+2\Re([\Qa,V]u,\Qa u)-4(\Bb V u,u) .
\end{gather}
To the first and the third terms in \eqref{6:Lower.6},
we apply \eqref{3:BP:equation.V}, and to the second
term we apply \eqref{3:BP:equation.VQ} (with $U=-2i\pd
V=[\Qa,V]$), obtaining
\begin{gather}\label{6:Lower.7}
2\Re(Vu,\PF_-u)\\
\nonumber =2(\Xc_{q+1}[\bb,V]v,v)-4(\Xc_{q}[\bb,\Bb V
]v,v)+4\Im(\Yc_q[\bb, \pd V]v,v).
\end{gather}
Finally, for the term $(\PF_-u,\PF_-u)$ we have
\begin{gather}\label{6:Lower.8}
\|\PF_-u\|^2=((\Q\Qa -2\Bb)u,(\Q\Qa
-2\Bb)u)=(\PF_-\Qa u,\Qa u) \\
\nonumber +4\Bn^2(u,u) +4((2\Bn+\bb)\bb u,u) - 4\Bn
\|\Qa u\|^2 -4\Re(\Qa\bb u,  \Qa u).
\end{gather}
All these  terms are of the form already considered in
\eqref{6:Lower.5},
\eqref{6:Lower.6},\eqref{6:Lower.7}. Thus we obtain
\begin{gather}\label{6:Lower.9}
(\PF_-u,\PF_-u)=C_{q+2}\|v\|^2+
(\Zc_{q+2}[\bb]v,v)-2\Bn C_{q+1}\|v\|^2\\
\nonumber -2\Bn(\Zc_{q+1}[\bb]v,v)
-2(\Xc_{q+1}[\bb,\bb]v,v)+4\Bn^2(C_q\|v\|^2+(\Zc_q[\bb]v,v))\\
\nonumber- 4\Bn (C_{q+1}\|v\|^2+(\Zc_{q+1}[\bb]v,v)) +
4(\Xc_q[\bb,(2\Bn+\bb)\bb
]v,v)\\
\nonumber-4(\Xc_{q+1}[\bb,\bb]v,v)-8\Im(\Yc_q[\bb,\pd
\bb]v,v).
\end{gather}

Now we collect the expressions \eqref{6:Lower.3},
\eqref{6:Lower.4}, \eqref{6:Lower.5},
\eqref{6:Lower.7} \eqref{6:Lower.9} to get
\begin{gather}\label{6:Lower.10}
\!\!\!\!\|(\PF_-\!+V-\!\m)u\|^2\! -\!
\t^2\|u\|^2\!=\!(\m^2-\t^2)(C_q\|v\|^2+(\Zc_q[\bb]v,v))\\
\nonumber
-2\m(\Xc_q[\bb,V]v,v)+(\Xc_q[\bb,V^2]v,v)-2\mu
C_{q+1}\|v\|^2-2\mu (\Zc_{q+1}[\bb]v,v)\\
\nonumber+4\mu\Bn
C_q\|v\|^2+4\mu\Bn(\Zc_{q}[\bb]v,v)+4\m
(\Xc_q[\bb,\bb]v,v)+2(\Xc_{q+1}[\bb,V]v,v)\\
\nonumber-4(\Xc_{q}[\bb,\bb V ]v,v)+4\Im(\Yc_q[\bb,
\pd V]v,v)+C_{q+2}\|v\|^2 +
(\Zc_{q+2}[\bb]v,v)\\
\nonumber-2\Bn C_{q+1}\|v\|^2 -2\Bn(\Zc_{q+1}[\bb]v,v)
-2(\Xc_{q+1}[\bb,\bb]v,v)+4\Bn^2C_q\|v\|^2\\
\nonumber+4\Bn^2(\Zc_q[\bb]v,v)- 4\Bn C_{q+1}\|v\|^2
-4\Bn(\Zc_{q+1}[\bb]v,v)\\\nonumber
 + 4(\Xc_q[\bb,(2\Bn+\bb)\bb
]v,v)-4(\Xc_{q+1}[\bb,\Bb]v,v)-8\Im(\Yc_q[\bb,\pd
\bb]v,v).
\end{gather}
The terms with $\|v\|^2$ have the coefficient
\begin{gather}\nonumber
    (\m^2-\t^2)C_q -2\mu
C_{q+1} +4\mu\Bn C_q     +C_{q+2}-2\Bn C_{q+1}
+4\Bn^2C_q   \\
\nonumber - 4\Bn
C_{q+1}=((\m^2-\t^2)-2\m\L+\L^2)C_q=\l(\L-s)C_q.
\end{gather}
The sum of all remaining terms in \eqref{6:Lower.10}
equals exactly $(-W_-v,v)$, where $W_-$ is given by
\eqref{6:Lower.00}. Now, by
Proposition~\ref{5:ToeplitzSubspaces} it is possible
to find a subspace $\Ms(\l)\subset\Hc_0$ such that
$(Wv,v)>\l(L-s)C_q\|v\|^2$ for $v\in \Hc_0$ and
dimension of $\Ms(\l)$ is greater than
$\X(\l(L-s)C_q)(1-o(\l))$. This is the subspace we
need.
\end{proof}
If the conditions of Theorem~\ref{6:Lower} are
satisfied then the conditions of Theorem~\ref{6:Upper}
are satisfied as well and we arrive at the asymptotic
formula
\begin{equation}\label{6:asymptotics}
\lim_{\l\to0}N_{\pm}(\l,q,\bb,V)\X(\l)^{-1}= 1
\end{equation}

\section{Proofs of technical lemmas}\label{technical}
\begin{proof}[Proof of Lemma~\ref{2:LemmaBound}] As usual,
 we may prove the inequality
for functions $u$ in the Schwartz space first, and
then extend it to all functions for which the right-hand
side of \eqref{2:LemmaBound.1} is finite, by continuity.

We will prove Lemma by a double induction.  For $N=1$
the statement is obvious: $||\Q u||^2=(\PF_-u,u),$
$||\Qa u||^2=(\PF_+u,u)=(\PF_-u,u)+2(\Bb u,u).$ Now,
suppose that for some $N_0$ we have established
\eqref{2:LemmaBound.1} for all $N<N_0$ and for all
combinations of $T_j$. The (exterior) induction step
will consist of proving \eqref{2:LemmaBound.1} for
$N=N_0$.

 To do this, we are going to show first that
one can replace $\PF_-$ by $\PF_+$ on the right-hand
side of \eqref{2:LemmaBound.1} for all $N<N_0$. This
will follow as soon as we prove  the inequality
 \begin{equation}\label{2:LemmaBound.1.1}
    (\PF_-^N u,u)\le C(\PF_+^N u,u)+C'(u,u),\; N\le N_0
\end{equation}
(note that the case $N=N_0$ \emph{is} included here.)
This fact we again prove by induction (interior) in
$N$. The statement is obvious for $N=1$. Suppose that
we have proved it for all $N<N_1\le N_0$ and we will
check it for $N=N_1$. In fact, by \eqref{2:ComRel.2},
\begin{equation}\label{2:LemmaBound.1.2}
\PF_-^{N_1}=\PF_+^{N_1} +\RF,
\end{equation}
 where $\RF$ is a sum of
many terms $\RF_\k$. Each $\RF_\k$ is the product of
less then $N_1$ factors $\PF_-$ and some factors
$2\Bb$. In other words, this means that $\RF_\k$ is a
product of no more than $N_1-1$ operators $\Q$, no
more than $N_1-1$ operators $\Qa$ and several
functions, with all derivatives bounded. Now we
transform each term $\RF_\k$ in the following way. We
move half of the factors $\Q,\Qa$ in $\RF_\k$ to the
left and the other half of these operators to the
right, commuting them with functions and using the
relations \eqref{3:BP:0}.

So, many terms will be produced, having the form
\begin{equation}\label{2:LemmaBound.2}
    T_1\dots T_k F T_{k+1}\dots T_{k+m},
\end{equation}
where $k,m<N_1$, each $T_j$ is $\Q$ or $\Qa$ and $F$
is a bounded function. Now we can estimate each term
in \eqref{2:LemmaBound.2}:
\begin{gather}\label{2:LemmaBound.3}
|(T_1\dots T_k F T_{k+1}\dots
T_{k+m}u,u)|=|(FT_{k+1}\dots T_{k+m}u, T_k^*\dots
T_1^* u)|\\ \notag \le C \|FT_{k+1}\dots
T_{k+m}u\|\;\|T_k^*\dots T_1^* u\|.
\end{gather}
The right-hand side in \eqref{2:LemmaBound.3} can be
estimated by the interior and the exterior inductive assumptions,
 since
there are less than $N_1$ and less than $N_0$ factors $T_j$
 involved.
Summing all these estimates over all terms generated
by $\RF$, we arrive at
 \begin{equation}\label{2:LemmaBound.4}\nonumber
    (\RF u, u)\le C(\PF_+^{N_1-1}u,u)+C'(u,u),
\end{equation}
which, together with \eqref{2:LemmaBound.1.2} leads to
\eqref{2:LemmaBound.1.1} for $N=N_1$.
Changing places of $\PF_+$ and $\PF_-$,
we obtain the dual  inequality
\begin{equation}\label{2:LemmaBound.5}
    (\PF_+^N u,u)\le C(\PF_-^N u,u)+C'(u,u),\; N\le
    N_0.
\end{equation}

Now, the inductive step in the exterior induction goes
rather easily. For $N=N_0$, let us have $N$ operators
$T_1,\dots, T_N$. Suppose that $T_N=\Q$. Then we can
write $||T_1\dots T_N u||^2$ as $||T_1\dots T_{N-1}
v||^2$ with $v=\Q u$. By the inductive assumption and
 \eqref{2:LemmaBound.1.1},
 we have $||T_1\dots T_{N-1}
v||^2\le C(\PF_+^{N-1} v,v)+C'(v,v)$,  the first term
here equals $C(\PF_-^Nu,u)$ so we arrive at
\eqref{2:LemmaBound.1}.
In the case of $T_N=\Qa$, we write the left-hand side
in \eqref{2:LemmaBound.1}
 as $||T_1\dots T_{N-1}
v||^2$
with $v=\Qa u$. By the inductive assumption, this
quantity is majorized by $(\PF_-^{N-1}v,v)+(v,v)$
which equals $(\PF_+^N u,u)+(\PF_+u,u)$. Applying
\eqref{2:LemmaBound.5}, we again obtain the
 estimate \eqref{2:LemmaBound.1}.
\end{proof}

\begin{proof}[Proof of Lemma~\ref{4:LemmaComp1}]
If $\AF$ satisfies the conditions of the Lemma and  we
commute some entry of the  resolvent with a
neighboring function or  the operators $\Q$ or $\Qa$
then all resulting operators will again satisfy these
conditions. So we perform such commutings by the
following rule. First, move one copy of the resolvent
to the utmost right position. Then if there are three
or more operators $\Q$, $\Qa$ in a row in front of
some resolvent, commute the utmost right of these
operator with resolvent. After performing this
operation as long as it is needed, we arrive at the
situation where in each summand  no more than two
operators $\Q,\Qa$ stand in front of each copy of the
resolvent. Each such group is a bounded operator, by
Lemma~\ref{2:LemmaBound}. At least one copy of the
resolvent has no more than one operator $\Q,\Qa$ in
front of it, because the order of the whole product is
negative. After this, we move the function $h$ that
has compact support to  the position in front of this
particular copy of the resolvent. In this process, the
structure of all products remains the same, moreover,
when we commute $h$ with resolvent, a compact term
arises. Finally, we arrive at the term $hT R_\pm$ or
$h R_\pm$, where $T$ is $\Q$ or $\Qa$. This operator
is, obviously, compact. All other factors in the
product are bounded, therefore the whole product is
compact.\end{proof}
\begin{proof}[Proof of Lemma~\ref{4:LemmaComp2}]
Boundedness of the operator in question is derived
from Lemma~\ref{2:LemmaBound} in the same way as
Proposition~\ref{2:PropComp} is proved. By means of
commuting, preserving the overall structure of all
arising terms, we arrive at a sum of operators of the
form

\begin{gather}\label{7:ProComp2.1}
\tilde{h}_\k\Tc_\k
P_q \Tc_\varkappa\tilde{h}_\varkappa =\\
\notag [\tilde{h}_\k \Tc_\k(\PF_-+1)^{-N}][(\PF_-+1)^N
P_q(\PF_-+1)^N]
[(\PF_-+1)^{-N}\Tc_\varkappa\tilde{h}_\varkappa],
\end{gather}
where $\Tc_\varkappa,\Tc_\k$ are products of no more
than $N$ operators $\Q, \Qa$. The middle term in
\eqref{7:ProComp2.1} is bounded by
Proposition~\ref{2:PropUnpertSubspace}, and the
extreme terms are bounded by Lemma~\ref{2:LemmaBound}.
If, say, the function $\tilde{h}_\k$ is compactly
supported, we can take $N$ larger than the number of
terms in $\Tc_\k$, and then the operator $\tilde{h}_\k
\Tc_\k(\PF_-+1)^{-N}$ has negative order and is
compact by Lemma~\ref{4:LemmaComp1}.\end{proof}


 \end{document}